\documentclass[useAMS,usegraphicx,usenatbib,letterpaper]{mn2e}
\usepackage{epsfig}
\usepackage{longtable}
\usepackage{graphics}
\usepackage{times}
\usepackage{epstopdf}
\usepackage[dvips]{graphicx}
\usepackage{latexsym}
\usepackage{amsmath}
\usepackage{amsfonts}
\usepackage{amssymb}
\usepackage{multicol}
\usepackage{array}
\usepackage{multirow}
\usepackage{subfigure}

\title[Coma Dwarf Galaxies: II. Fundamental Plane]{Dwarf Galaxies in the Coma Cluster: II. Spectroscopic and Photometric Fundamental Planes\thanks{Based in part on observations made with the NASA/ESA Hubble Space Telescope, obtained from the data archive at the Space Telescope Institute. STScI is operated by the association of Universities for Research in Astronomy, Inc. under the NASA contract  NAS 5-26555. These observations are associated with programme GO10861.}\thanks{Some of the data presented herein were obtained at the W.M. Keck Observatory, which is operated as a 
scientific partnership among the California Institute of Technology, the University of California and the 
National Aeronautics and Space Administration. The Observatory was made possible by the generous 
financial support of the W.M. Keck Foundation. }}
\author[Kourkchi et al.]
       {E. Kourkchi$^{1,2}$, H. G. Khosroshahi$^{1}$, D. Carter$^{3}$, B. Mobasher$^{4}$
      \\$^1$ School of Astronomy, Institute for Research in Fundamental Sciences (IPM), PO Box 19395-5531,Tehran, Iran.
       \\$^2$ Department of Physics, Sharif University of Technology, P.O.Box:11155-9161, Tehran, Iran.
      \\$^3$ Astrophysics Research Institute, Liverpool John Moores University, Twelve Quays House, Egerton
      Wharf, Birkenhead CH41 1LD, UK.
      \\$^4$Department of Physics and Astronomy, University of California, Riverside, CA 92521, USA.
      }

\begin{document}

\date{Accepted 2011 October 10.  Received 2011 September 13; in original form 2011 February 4}
\pagerange{\pageref{firstpage}--\pageref{lastpage}}
\pubyear{2011} \volume{000} \pagerange{1}

\maketitle 

\label{firstpage}

\begin{abstract}

We present a study of the fundamental plane, FP, for a sample of 71 dwarf galaxies in the core of Coma cluster in magnitude range $-21 < M_I <-15$. Taking advantage of high resolution DEIMOS spectrograph on Keck II for measuring the internal velocity dispersion of galaxies and high resolution imaging of HST/ACS, which allows an accurate surface brightness modeling, we extend the fundamental plane (FP) of galaxies to $\sim$1 magnitude fainter luminosities than all the previous studies of the FP in Coma cluster. We find that, the scatter about the FP depends on the faint-end luminosity cutoff, such that the scatter increases for fainter galaxies. The residual from the FP correlates with the galaxy colour, with bluer galaxies showing larger residuals from FP.

We find $M/L \propto M^{-0.15\pm0.22}$ in F814W-band indicating that in faint dwarf ellipticals, the $M/L$ ratio is insensitive to the mass. We find that less massive dwarf ellipticals are bluer than their brighter counterparts, possibly indicating ongoing star formation activity. Although tidal encounters and harassment can play a part in removing stars and dark matter from the galaxy, we believe that the dominant effect will be the stellar wind associated with the star formation, which will remove material from the galaxy resulting in larger $M/L$ ratios. We attribute the deviation of a number of  faint blue dwarfs from the FP of brighter ellipticals to this effect.

We also study other scaling relations involving galaxy photometric properties including the photometric plane. We show that, compared to the FP, the scatter about the photometric plane is smaller at the faint end.

\end{abstract}


\section{Introduction}
\label{chap:introduction}
\label{introduction}

The photometric and dynamical properties of the elliptical galaxies could be explained by multivariate analysis which reveals the fundamental scaling relations of these galaxies. One of the most well-known manifolds showing a tight correlation between $R_e$ (i.e. the radius encompassing half-light of the galaxy), $I_e$ (i.e. the mean surface brightness within $R_e$ in flux unit), and $\sigma$ (i.e. the galaxy internal velocity dispersion), is the Fundamental Plane (hereafter FP; Djorgovski \& Davis 1987; Faber et al. 1987, Dressler et al. 1987; Bernardi et al. 2003).

In theory, the FP is derived from the virial theorem as $R_e\propto \sigma^2 I_e^{-1} (M/L)^{-1}$, where $I_e$ is the effective surface brightness in flux units, calculated within the half-light radius, $R_e$, of the galaxy, $\sigma$ is the galaxy internal velocity dispersion and $M/L$ is its mass-to-light ratio. Assuming that the $M/L$ is expressed by a power-law function of $\sigma$ and/or the effective surface brightness, $I_e$, the FP relation is simplified as

\begin{equation}
log(R_e)=A~log(\sigma)+B ~ \langle \mu \rangle_e+C ,
\label{fp}
\end{equation}

where $\langle \mu \rangle_e$ is the mean surface brightness in $mag/arcsec^2$ unit and is defined as $\langle \mu \rangle_e=-2.5log(I_e)+cte$. Although the shape of the FP and its coefficients differs for different gravitationally bound systems from globular clusters (Burstein et al. 1997) to galaxy clusters (Schaeffer et al. 1993; Fritsch \& Buchert 1999; Zaritsky et al. 2006a: ZGZ06), there is no doubt about its existence (Lucey, Bower \& Ellis 1991 and its references). In reality, the coefficients of the FP relation differ from the prediction of the virial theorem. The observed coefficients are A=1.24 and B=0.33 (J{\o}rgensen, Franks and Kj\ae rgaard 1996: JFK96) while the virial theorem predicts A=2.0 and B=0.4. This difference, often referred to as the ``tilt" of the FP, is mainly attributed to different formation histories and evolutionary processes.

\begin{figure}
\begin{center}
\includegraphics[width=8cm]{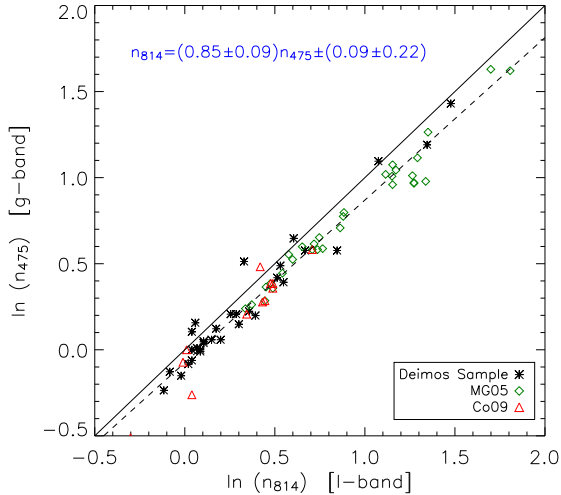}
\caption {Comparing the S\'{e}rsic parameters in two different filters (F814W \& F475W). For both filters, the initial parameters to run Galfit are chosen from the SExtractor catalogue of F814W images. The solid black line is the locus of $n_{814}=n_{475}$. The blue dashed line is the best linear regression between the two parameters. $n_{814}$ values are 10\% higher than $n_{475}$ values, on average. Red triangles and green diamonds represent galaxies of Co09 and MG05 samples, respectively. Black asterisks are dwarf galaxies of our DEIMOS sample. }
\label{fig:ser81475}
\end{center}
\end{figure}

The difference in FP coefficients of different spheroidal systems with different mass, size and luminosities, can be explained by evolution of the $M/L$ ratio as a function of stellar population [age, metalicity or initial mass function (IMF)] and/or dark matter content (Tortora et al. 2009; Grillo \& Gobat 2010; Graves \& Faber 2010). In addition, the absence of homology, i.e, the fact that, the structure of spheroids is scalable regardless of their size, can be the source of the FP tilt (D'Onofrio et al. 2008; Trujillo, Burkert \& Bell 2004). On the other hand, some authors studied the role of dissipation in explaining the nature of the FP (Ribeiro \& Dantas 2010; Hopkins, Thoms \& Hernquist 2008: HCH08). HCH08 claimed that the non-homology or change in the dark matter distribution are not the main drivers of FP tilt. Studying the early-type galaxies in 59 nearby galaxy clusters, D'Onofrio et al. (2008) have found a strong correlation between the FP coefficients and the local cluster environment and no strong correlations with internal galaxy properties (e.g. S'ersic index and galaxy colour). Moreover, FP coefficients are independent of global properties of clusters such as radius, X-ray luminosity and cluster velocity dispersion (D'Onofrio et al., 2008). On the other hand, while Reda, Forbes \& Hau (2005) have shown that isolated early-type galaxies lie on the same fundamental plane as galaxies in high-density environments, cluster galaxies have also less intrinsic scatter in their properties compared to field galaxies (de Carvalho \& Djorgovski, 1992).

\begin{figure*}
\centering

{
   \label{fig:sub:a}
   \includegraphics[width=2.3in]{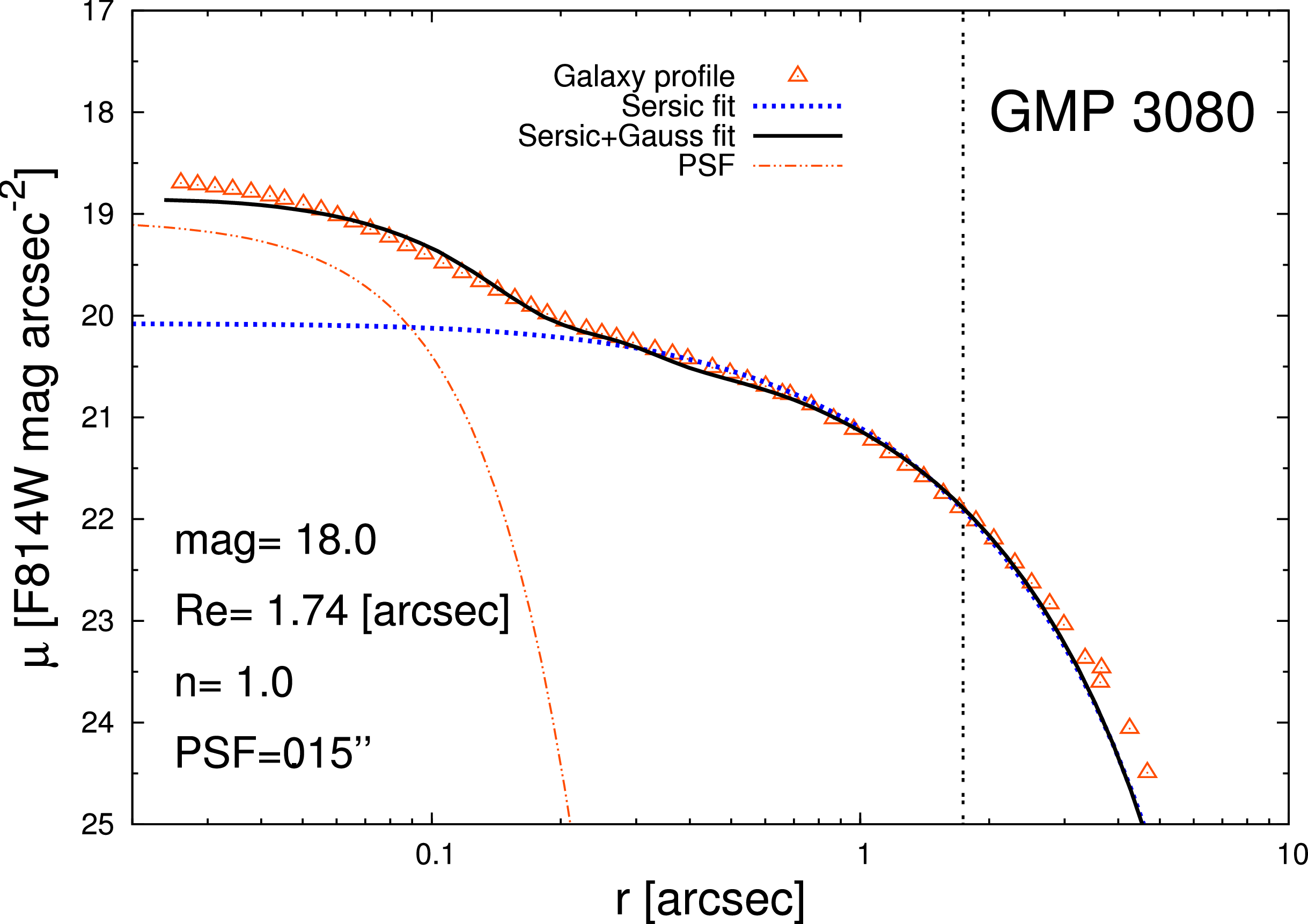}
}
\hspace{0.8in}
{
   \label{fig:sub:b}
   \includegraphics[width=2.55in]{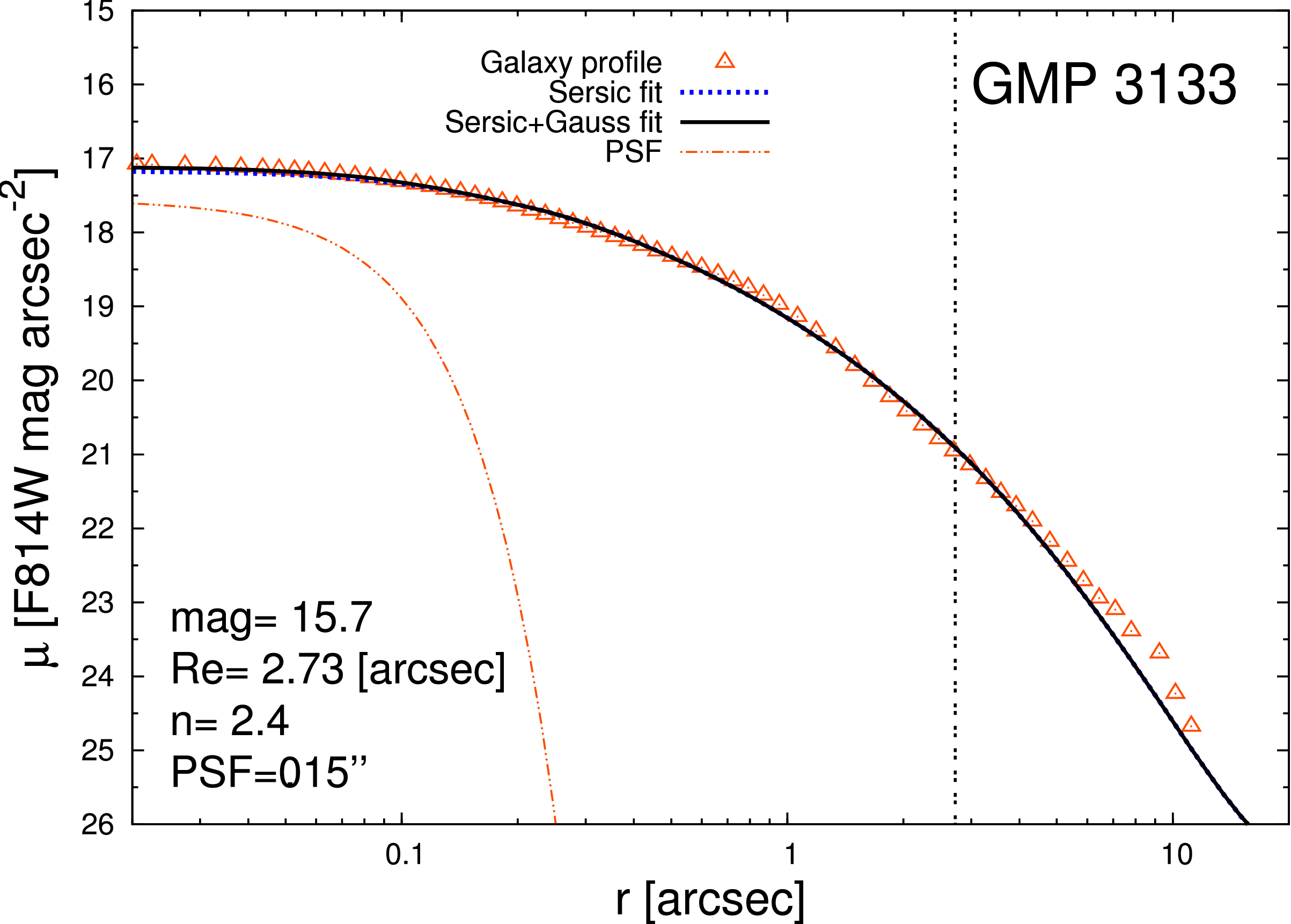}

\vspace{0.2in}
}
\caption{Two examples of the fitted S\'{e}rsic function to the galaxy light profile. Red open triangles show the real profile of the galaxies while the blue dotted and black solid curves represent the best fit from a single S\'{e}rsic and S\'{e}rsic+gaussian models, respectively. The vertical black double-dotted lines show the location of the effective radius, $R_e$, of these galaxies. The horizontal axes show the average radius of the ellipsoidal isophotes ($r=\sqrt{ab}$). The red dash-dotted curves represent the typical profile of the PSF. {\bf Left)} It shows the surface brightness profile of {\it GMP3080}. The central excess light forced us to add an extra gaussian component which results in the best fit with the S\'ersic index of n=1.0 (the black solid curve). The FWHM of the added central Gaussian function is $0.13"$. The S\'{e}rsic index corresponding to the blue dotted curve is 0.9. {\bf Right)} The surface brightness profile of {\it GMP3133}, which is accurately described with a single S\'ersic profile.}
\label{fig:galfit} 
\end{figure*}

The study of dark matter in dwarf galaxies showed that below the critical virial velocity, which is estimated to be $\sim$100 km s$^{-1}$, interstellar gas removal via supernova explosions become important (Dekel \& Silk 1986: DS86). This mechanism has been invoked also to explain the shape of the low mass dwarf galaxies and their mass profiles (S\'anchez-Janssen et al. 2010; Governato et al. 2010). Differences between the FP of giant and dwarf galaxies have been known for some time (Nieto et al. 1990; Bender et al. 1992; Guzman et al. 1993). Peterson \& Caldwell (1993), studying a sample of nucleated dwarfs, they found a change in $M/L$ ratio with luminosity as predicted by the scaling relations of DS86. The study of 17 Virgo dwarfs (-17.5$<M_V<$-15.48) by Geha et al. (2002 \& 2003) placed dEs on a plane parallel to, but offset, from that occupied by normal elliptical galaxies. On the other hand, Graham \& Guzm\'an (2003), based on a detailed surface photometry from HST archival images for a sample of dEs in the Coma cluster, claimed that the E-dE dichotomy is a consequence of a continuously varying profile shape with galaxy luminosity. They attributed the E-dE dichotomy to the fact that the light profile of dwarf galaxies do not follow the de Vaucouleurs law (1948) and therefore, concluded that giant and dwarf ellipticals (hereafter dEs) were not formed by different mechanisms. However, as found by Geha et al. (2003), modelling the dEs with S\'ersic functions (1968) does not explain why dEs lie on a different region of the FP. Although it is clear that dwarf ellipticals do not lie on the fundamental plane as defined by brighter ellipticals (Geha et al. 2003), Zaritsky et al. (2006 a,b) showed that the differences between the ``fundamental planes'', defined by different families of spheroids from dwarf ellipticals to galaxy clusters, is the result of a non-linear relationship between $\log{\sigma}$ and $\log{M/L}$. Desroches et al. (2007) also found the sensitivity of the coefficient A of the FP to the luminosity faint-end cutoff. D'Onofrio et al. (2008) were among the first to question the universality of FP and proposed the bending of FP of early-type galaxies. On the other hand, Fraix-Burnet et al. (2010) studied a sample of galaxies with the redshift range 0.007-0.053, and found that the global FP is a mixture of several fundamental planes, with different thickness and orientations, and concluded that the FP is not a bent surface. They found seven groups of fundamental planes which are assigned to different galaxy assembly histories and formation scenarios.

Another fundamental scaling relation, based on pure photometric observables, was introduced by Khosroshahi et al (2000 a,b), called the Photometric Plane (hereafter PHP) of the galaxies. The S\'{e}rsic index in the PHP, which replaces the velocity dispersion in FP, carries information on the light profile in galaxies and it is correlated with the dynamical properties of the galaxies such as the total mass and internal stellar velocity dispersion (Graham 2002: GR02). Steeper the light profile at the centre of the galaxy (higher S\'ersic index), more massive the galaxy and larger the central velocity dispersion.

FP and PHP could be potentially used to derive distance to any single galaxy. GR02 has compared both photometric and fundamental planes for exactly the same sample of early-type galaxies in Virgo and Fornax clusters, and concluded that the scatter about the PHP is $\sim$15-30\% more than the scatter about FP. Since the PHP is purely based on photometric quantities, it can be replaced with FP wherever the expensive kinematic data are not available and/or high level of accuracy is not required.

In Paper I in this series (Kourkchi et al. 2011a) we used the high spectral resolution of DEIMOS (Faber et al. 2003) on the Keck II telescope to measure precise velocity dispersions for a sample of 41 faint elliptical galaxies in Coma cluster. Of this sample, 28 galaxies lie in the observed footprints of the HST/ACS Coma Treasury survey (Carter et al. 2008). To this sample we add a further 43 galaxies with velocity dispersions from Matkovi\'c \& Guzm\'an (2005: MG05) or Cody et al. (2009: Co09), which also lie in these footprints, giving us a combined sample of 71 objects with $-22 < M_R < -15$ with precise velocity dispersions measured from one of these sources. We noticed that fainter dwarfs show a departure from brighter ellipticals on the Faber Jackson relation (Faber \& Jackson 1976) which indicates they have higher velocity dispersion and therefore are more massive or less luminous than is predicted by L-$\sigma$ linear trend.

Any study of the differences in the FP of dwarf and giant ellipticals helps us to test the galaxy formation scenarios. In this paper, we present the structural parameters (effective radius, effective surface brightness, central surface brightness, S\'ersic index and concentration parameter) for galaxies in our sample, based on the Treasury HST/ACS images in the F814W and F475W filters. The precise measured structural parameters from the HST/ACS images helps to examine the form of the scaling relations at faint luminosity and low surface brightness, and to investigate how the Coma dwarfs fit onto the well-known fundamental plane. In addition, we study the scatter of faint dEs about the fundamental plane of giant ellipticals and examine whether it depends upon photometric and structural parameters of galaxies, such as S\'ersic index, luminosity and colour. Moreover, we examine the form of the photometric plane of our sample dEs.

This paper is arranged as follows. \S \ref{chap:phanalysis} describes the photometric data analysis and modelling the surface brightness of galaxies. \S \ref{chap:scales} is devoted to study the scaling relations of our sample galaxies, such as the fundamental plane, photometric plane, size-surface brightness, and size-luminosity relations, as well as the correlation between different photometric and kinematic parameters. Departure of faint dEs from the FP of bright elliptical is discussed in \S \ref{chap:departure}. In \S \ref{mtolight}, we study the variation of the $M/L$ ratio of the galaxies in our sample across their total mass. Discussion and summary are drawn in sections \S \ref{discussion} and \S \ref{conclusions}.

Throughout this paper, we assume that the Coma cluster is located at a distance of 100 Mpc (z=0.0231), which corresponds to a distance modulus of
35.00 mag and angular scale of 0.463 kpc arcsec$^{-1}$ for $h=71$, $\Omega_m=0.27$ and $\Omega_\lambda=0.73$ (Carter et al. 2008). In this paper,
all magnitudes are in the AB system.



\begin{table*}
\label{tab:results}
\caption{The photometric and dynamical parameters of early-type galaxies in Coma cluster of which 28 galaxies have spectroscopic data from DEIMOS spectrograph. The velocity dispersion for 34 galaxies were extracted from MG05 and for 9 galaxies from Co09. The photometric parameter are derived from ACS images taken by filter F814W. $\mu^*_0$ is the central surface brightness of each galaxy found by the IRAF task {\it ellipse} and $\mu_0$ is the same but for galaxy model created by {\it Galfit}. $R_e$ is the effective radius of the galaxy encloses half of the total light of the galaxy. $n$ and $\sigma$ (column 10 and 11) are the S\'{e}rsic index and velocity dispersion respectively. C (column 12) is the concentration parameter which is defined as $C=5~log_{10}(r_{80}/r_{20})$.}

\centering
\begin{tabular}[tc]{ c  c c c | c  c  c  c |  c  c  c  c }

\hline 

\hline
GMP & RA & DEC & mag (AB)  & $\mu^*_0$ &  $\mu_0$ &  $\mu_e$ & $\langle \mu \rangle_e$ & $R_e $ & $n$  &  $\sigma$  & $C$   \\  
$ID$ & J2000 & J2000 & $F814W$ &  \multicolumn{4}{c|}{{$mag/arcsec^{2}$}}  &  $kpc$ &  &  $kms^{-1}$ &     \\
(1) & (2) & (3) & (4) &  (5) & (6) &  (7) & (8) &  (9) &  (10) &  (11) &  (12)  \\ \hline \hline

2489 & 13:00:44.69 & 28:06:01.00 & 15.1 & 16 & 16.5 & 20.7 & 19.6 & 1.52 & 2.1 &          94$\pm$2          &  4.11 \\ 
2510 & 13:00:42.92 & 27:57:45.49 & 14.8 & 15.5 & 15.5 & 22.5 & 20.8 & 4.47 & 6.1 &          127$\pm$3          &  4.22 \\ 
2516 & 13:00:42.85 & 27:58:14.90 & 14.1 & 15 & 14.9 & 21.3 & 19.7 & 3.72 & 5.5 &          176$\pm$5          &  3.86 \\ 
2529 & 13:00:41.28 & 28:02:41.00 & 16.9 & 18.1 & 18.6 & 20.9 & 20 & 0.78 & 1.4 &          42$\pm$6          &  3.24 \\ 
2535 & 13:00:40.94 & 27:59:46.19 & 14.6 & 15.8 & 16.3 & 21.2 & 20 & 2.79 & 3.1 &          124$\pm$2          &  3.97 \\ 
2541 & 13:00:39.83 & 27:55:24.40 & 14.1 & 15.4 & 15.6 & 20.8 & 19.4 & 2.88 & 3.6 &          177$\pm$4          &  3.82 \\ 
2563 & 13:00:37.30 & 27:54:41.10 & 19.4 & 19.1 & 21.4 & 23 & 22.3 & 0.66 & 1.1 &          25$\pm$10          &  2.84 \\ 
2571 & 13:00:36.60 & 27:55:52.00 & 18.5 & 19.2 & 20.2 & 22.5 & 21.7 & 0.75 & 1.3 &          17$\pm$6          &  3.05 \\ 
2585 & 13:00:35.50 & 27:56:32.15 & 17.1 & 19.2 & 19.4 & 22.6 & 21.7 & 1.64 & 1.6 &          30$\pm$6          &  3.34 \\ 
2591 & 13:00:34.40 & 27:56:05.00 & 17.2 & 19.1 & 19.7 & 22.5 & 21.6 & 1.56 & 1.7 &          53$\pm$9          &  3.18 \\ 
2605 & 13:00:33.30 & 27:58:49.00 & 18.3 & 19.6 & 20.9 & 22.9 & 22.2 & 1.05 & 1.2 &          36$\pm$10          &  2.89 \\ 
2654 & 13:00:28.00 & 27:57:22.00 & 14.8 & 15 & 15.3 & 20.4 & 19 & 1.38 & 4.4 &          144$\pm$7          &  4.76 \\ 
2655 & 13:00:27.90 & 27:59:16.50 & 19.2 & 20.4 & 21.1 & 23 & 22.3 & 0.71 & 1.1 &          45$\pm$5          &  2.84 \\ 
2676 & 13:00:26.20 & 28:00:32.00 & 17.8 & 18.7 & 20.2 & 22.1 & 21.4 & 0.93 & 1.1 &          37$\pm$8          &  2.94 \\ 
2692 & 13:00:24.90 & 27:55:34.14 & 16.8 & 18.4 & 19.3 & 23 & 22 & 2.16 & 1.6 &          41$\pm$6         &  3.58 \\ 
2718 & 13:00:22.70 & 27:57:55.00 & 18.6 & 19.9 & 20.7 & 22.8 & 22.1 & 0.88 & 1.3 &          30$\pm$8          &  3.01 \\ 
2736 & 13:00:21.70 & 27:53:55.00 & 16.7 & 17.4 & 18.5 & 21.4 & 20.4 & 1.08 & 1.7 &          35$\pm$4          &  3.43 \\ 
2755 & 13:00:20.20 & 27:59:38.00 & 18.9 & 20.4 & 20.9 & 22.5 & 21.9 & 0.69 & 1 &          27$\pm$9          &  2.8 \\ 
2778 & 13:00:18.88 & 27:56:11.75 & 15.6 & 17.7 & 18.2 & 21.9 & 20.8 & 2.47 & 2.1 &          57$\pm$4          &  3.36 \\ 
2780 & 13:00:18.70 & 27:55:13.00 & 19 & 19.5 & 20.7 & 22.9 & 22.1 & 0.69 & 1.4 &          63$\pm$12          &  3.02 \\ 
2784 & 13:00:18.62 & 28:05:48.00 & 17 & 19 & 19.4 & 21.9 & 21.1 & 1.24 & 1.5 &          64$\pm$8          &  3.12 \\ 
2799 & 13:00:17.72 & 27:59:13.29 & 17.7 & 18.2 & 18.4 & 21.3 & 20.4 & 0.65 & 2.2 &          45$\pm$9          &  3.51 \\ 
2805 & 13:00:17.11 & 28:03:48.00 & 15.1 & 15.1 & 15.7 & 20.2 & 19 & 1.22 & 3.6 &          129$\pm$3          &  4.21 \\ 
2808 & 13:00:17.00 & 27:54:16.10 & 19.6 & 19.9 & 20.4 & 22.1 & 21.5 & 0.43 & 0.9 &          69$\pm$12          &  2.85 \\ 
2839 & 13:00:14.83 & 28:02:27.00 & 14.4 & 14.5 & 15.1 & 19.7 & 18.3 & 1.24 & 3.8 &          168$\pm$6          &  4.41 \\ 
2861 & 13:00:12.98 & 28:04:30.00 & 14.8 & 15.4 & 15.6 & 20.1 & 18.9 & 1.37 & 3.2 &          123$\pm$2          &  4.03 \\ 
2877 & 13:00:11.40 & 27:54:36.00 & 18.6 & 19.3 & 20.2 & 23.1 & 22.2 & 0.96 & 1.8 &          29$\pm$9          &  3.23 \\ 
2879 & 13:00:11.23 & 28:03:53.00 & 16.7 & 17.9 & 18.8 & 21.4 & 20.5 & 1.1 & 1.6 &          44$\pm$5          &  3.19 \\ 
2922 & 13:00:08.10 & 28:04:41.00 & 14.3 & 15.1 & 15.6 & 18.7 & 17.7 & 0.93 & 2.1 &          185$\pm$4          &  3.65 \\ 
2931 & 13:00:07.10 & 27:55:52.00 & 17.6 & 17.2 & 17.9 & 20.7 & 19.6 & 0.46 & 2.3 &          35$\pm$6          &  3.71 \\ 
2960 & 13:00:05.47 & 28:01:26.00 & 15.5 & 17.6 & 18 & 21.8 & 20.8 & 2.37 & 2 &          60$\pm$5          &  3.69 \\ 
3017 & 13:00:01.05 & 27:56:41.80 & 16.9 & 17.8 & 18.4 & 21.9 & 20.9 & 1.22 & 2.4 &          59$\pm$12          &  3.69 \\ 
3018 & 13:00:01.00 & 27:59:30.00 & 18 & 18.1 & 20.2 & 22 & 21.3 & 0.81 & 1.1 &          38$\pm$5          &  2.95 \\ 
3034 & 12:59:59.56 & 27:56:24.42 & 17.1 & 18.4 & 19.6 & 22.7 & 21.9 & 1.98 & 1.9 &          170$\pm$84          &  3.15 \\
\hline \multicolumn{12}{r}{{Continued on next page}} \\
\end{tabular}
\end{table*}

\setcounter{table}{0}
\begin{table*}
\label{tab:results}
\caption{Continued from previous page}

\centering
\begin{tabular}[tc]{ c  c c c | c  c  c  c |  c  c  c  c }

\hline
GMP & RA & DEC & mag (AB)  & $\mu^*_0$ &  $\mu_0$ &  $\mu_e$ & $\langle \mu \rangle_e$ & $R_e $ & $n$  &  $\sigma$  & $C$   \\  
$ID$ & J2000 & J2000 & $F814W$ &  \multicolumn{4}{c|}{{$mag/arcsec^{2}$}}  &  $kpc$ &  &  $kms^{-1}$ &     \\
(1) & (2) & (3) & (4) &  (5) & (6) &  (7) & (8) &  (9) &  (10) &  (11) &  (12)  \\ \hline \hline

3068 & 12:59:56.75 & 27:55:46.40 & 15.1 & 16.1 & 18 & 21.4 & 20.4 & 2.44 & 1.8 &          106$\pm$2          &  3.5 \\ 
3080 & 12:59:55.70 & 27:55:04.00 & 17.9 & 18.8 & 20.1 & 21.9 & 21.1 & 0.79 & 1.1 &          9$\pm$8          &  2.96 \\ 
3098 & 12:59:53.90 & 27:58:14.00 & 17.6 & 18.9 & 20.3 & 21.8 & 21.1 & 0.93 & 0.9 &          33$\pm$7          &  2.76 \\ 
3119 & 12:59:51.50 & 27:59:35.50 & 19.6 & 19.9 & 21.5 & 23.6 & 22.6 & 0.65 & 1.2 &          37$\pm$10          &  2.79 \\ 
3131 & 12:59:50.20 & 27:54:46.00 & 17.3 & 18.9 & 20.2 & 22.1 & 21.4 & 1.28 & 1 &          19$\pm$7          &  2.79 \\ 
3133 & 12:59:50.18 & 27:55:27.65 & 15.9 & 17.1 & 17.2 & 20.9 & 19.9 & 1.26 & 2.4 &          80$\pm$3          &  3.61 \\ 
3141 & 12:59:49.10 & 27:58:33.90 & 19.7 & 19.4 & 21.5 & 23.2 & 22.5 & 0.58 & 1 &          59$\pm$15          &  2.88 \\ 
3146 & 12:59:48.60 & 27:58:58.00 & 17.8 & 20.3 & 20.9 & 22.8 & 22 & 1.35 & 1 &          38$\pm$11          &  2.78 \\ 
3166 & 12:59:46.90 & 27:59:31.00 & 17.2 & 19.3 & 20 & 22 & 21.3 & 1.16 & 1.1 &          38$\pm$6          &  3 \\ 
3170 & 12:59:46.88 & 27:58:24.04 & 14.5 & 15.6 & 16.1 & 20.9 & 19.6 & 2.35 & 3.2 &          143$\pm$4          &  4.3 \\ 
3201 & 12:59:44.47 & 27:54:43.02 & 14.1 & 15.3 & 16 & 20 & 19 & 2 & 2.4 &          170$\pm$4          &  3.7 \\ 
3209 & 12:59:44.20 & 28:00:47.00 & 18.5 & 19.4 & 20.2 & 22 & 21.3 & 0.66 & 1 &          30$\pm$5          &  2.86 \\ 
3213 & 12:59:43.80 & 27:59:39.15 & 14.8 & 15.4 & 15.8 & 19.8 & 18.6 & 1.15 & 3.2 &          139$\pm$5          &  4.01 \\ 
3222 & 12:59:42.38 & 27:55:27.37 & 15 & 14 & 14.2 & 19.8 & 18.1 & 0.91 & 7.6 &          164$\pm$4          &  4.9 \\ 
3223 & 12:59:42.40 & 28:01:58.60 & 18.7 & 19.8 & 21.2 & 23.5 & 22.6 & 1.19 & 1.4 &          22$\pm$9          &  2.62 \\ 
3254 & 12:59:40.36 & 27:58:03.95 & 15.6 & 15.5 & 16.3 & 21 & 19.9 & 1.45 & 3.6 &          101$\pm$5          &  4.18 \\ 
3269 & 12:59:39.73 & 27:57:12.57 & 15.3 & 15.4 & 15.9 & 20.5 & 19.3 & 1.3 & 3.2 &          99$\pm$3          &  4.18 \\ 
3292 & 12:59:38.00 & 28:00:03.70 & 16.6 & 17.1 & 18.8 & 21.4 & 20.5 & 1.12 & 1.5 &          40$\pm$5          &  3.21 \\ 
3308 & 12:59:37.20 & 27:58:20.00 & 18 & 16.3 & 16.6 & 18.7 & 17.6 & 0.16 & 2.9 &          50$\pm$4          &  3.87 \\ 
3312 & 12:59:37.00 & 28:01:07.00 & 17.5 & 17.9 & 18.3 & 20.8 & 19.9 & 0.56 & 1.7 &          31$\pm$4          &  3.38 \\ 
3339 & 12:59:35.37 & 27:51:47.40 & 16.1 & 17.8 & 17.9 & 20.7 & 19.8 & 1.08 & 1.7 &          59$\pm$4          &  3.17 \\ 
3367 & 12:59:32.86 & 27:58:59.55 & 14.1 & 14.7 & 15.7 & 20.9 & 19.5 & 2.84 & 3.5 &          179$\pm$4          &  4.17 \\ 
3400 & 12:59:30.90 & 27:53:01.74 & 14 & 14.5 & 14.6 & 19.4 & 18.1 & 1.42 & 3.9 &          217$\pm$5          &  3.86 \\ 
3406 & 12:59:30.30 & 28:01:15.10 & 17.7 & 18.5 & 19.2 & 21.6 & 20.8 & 0.76 & 1.4 &          36$\pm$3          &  3.22 \\ 
3438 & 12:59:28.50 & 28:01:09.40 & 18 & 18.8 & 20.1 & 21.9 & 21.2 & 0.78 & 1.2 &          26$\pm$8          &  3 \\ 
3473 & 12:59:26.40 & 27:51:25.00 & 17.6 & 18.8 & 19.2 & 21.6 & 20.7 & 0.78 & 1.5 &          33$\pm$4          &  3.27 \\ 
3681 & 12:59:11.64 & 28:00:31.00 & 16.6 & 18.1 & 18.5 & 21.1 & 20.3 & 1.01 & 1.6 &          64$\pm$5          &  3.31 \\ 
3707 & 12:59:09.53 & 28:02:26.00 & 16.2 & 17.5 & 17.8 & 21.7 & 20.6 & 1.49 & 2 &          77$\pm$4          &  3.77 \\ 
3780 & 12:59:04.87 & 28:03:00.00 & 16.4 & 18.1 & 18.6 & 22.1 & 21.2 & 1.75 & 1.8 &          58$\pm$3          &  3.54 \\ 
4035 & 12:58:45.50 & 27:45:14.00 & 17.1 & 19.1 & 19.7 & 22.3 & 21.4 & 1.49 & 1.6 &          26$\pm$4          &  3.14 \\ 
4135 & 12:58:37.30 & 27:10:35.00 & 15.1 & 18.3 & 19.1 & 20.6 & 20.1 & 1.93 & 0.7 &          42$\pm$3          &  2.55 \\ 
4215 & 12:58:31.70 & 27:23:42.00 & 17.6 & 20.5 & 20.7 & 23.3 & 22.6 & 1.82 & 1 &          7$\pm$4           &  3.11 \\ 
4381 & 12:58:15.30 & 27:27:53.00 & 17.7 & 18.4 & 18.6 & 21.8 & 20.8 & 0.81 & 2 &          22$\pm$4          &  3.56 \\ 
4430 & 12:58:20.50 & 27:25:46.00 & 16.6 & 18.8 & 18.9 & 21.9 & 21 & 1.49 & 1.4 &          38$\pm$3          &  3.17 \\ 
5102 & 12:57:04.30 & 27:31:34.00 & 15.9 & 18.1 & 19.2 & 21.9 & 21.2 & 2.18 & 1 &          42$\pm$3          &  3.18 \\ 
5364 & 12:58:33.10 & 27:21:52.00 & 15 & 16.6 & 18.5 & 21.8 & 20.9 & 3.26 & 1.5 &          84$\pm$4          &  4.11 \\ 
5365 & 12:56:34.60 & 27:13:40.00 & 15.5 & 17.6 & 19.4 & 21.4 & 20.7 & 2.37 & 1 &          44$\pm$3          &  3.01 \\
\hline
\end{tabular}
\end{table*}

\section{The Photometric Data}
\label{chap:phanalysis}

The HST ACS Coma cluster treasury survey is a deep two-passband imaging survey of one of the nearest rich cluster of galaxies. The completed survey covers 274 arcmin$^2$ area of sky in the core and infall region of the Coma cluster. 25 fields were imaged by ACS Wide Field Camera with the F475W (g-band) and F475W (I-band) filters. Of 25 fields, 19 were located within 0.5 Mpc (0.3 deg) of Coma centre. For the purpose of this study, we used the images of Data Release 2 (DR2), which include several improvements to the initial release. 

\subsection{Photometry of Sample Galaxies}
\label{chap:photoanalysis}

Among the galaxies with DEIMOS spectroscopic data, 32 have HST/ACS images. Reliable velocity dispersions were derived for 28 of these galaxies. Four remaining galaxies were therefore excluded from the analysis. Moreover, velocity dispersion measurements were available for 41 more galaxies (34 from MG05 and 9 from Co09) in Coma for which ACS images at the above bands were available. The final sample consists of 71 galaxies with ACS images and determined velocity dispersion covering a luminosity range from $M_R\approx$ -22 to -15.

For the purpose of this study, galaxies in both F814W/F475W images were extracted using {\it SExtractor} (Bertin \& Arnouts 1996) for photometry and measuring the initial model independent shape parameters such as the effective radius, $R_e$, position angle and ellipticity. For each galaxy, the initial central position as well as the concentration parameter (i.e. $C=5~log_{10}({r_{80}/r_{20})}$, where $r_{80}$ ($r_{20}$) is the radius within which $\%80$ ($\%20$) of the total galaxy light is collected) and the Kron radius, i.e. a characteristic radius as weighted by the light profile originally defined by Kron (1980), were measured using SExtractor. For each galaxy, we used both F814W/F475W bands to run SExtractor in dual-image mode where F814W-band was used for object detection. The SExtractor input parameters are directly taken from Table 1 of Hammer et al. 2010.

\subsection{Galaxy Surface Brightness Profile}
\label{chap:structure}

S\'ersic function, defined as log I(r)$\propto$r$^{1/n}$, describes the structure of most elliptical galaxies remarkably well (Kormendy, 2009). To find the best S\'{e}rsic fit to the light profile, we used {\it Galfit} (version 3, Peng et al. 2010). For ACS images, the anisotropic PSF shape depends on the location of each object on ACS CCD chips (WFC1 \& WFC2) and was modelled by {\it tinytim} (Krist, 1993). To run Galfit, initial values of the $R_e$, $\mu_e$, galaxy position and galaxy position angle were taken from SExtractor initial run, and the initial value of the S\'ersic index $n$ was set to 3. Tests show that with the well defined psf of ACS data the final solution does not depend strongly upon the initial value of $n$ (see also Hoyos et al. 2011). Any object in the vicinity of the target galaxies was masked out. It is important to leave enough sky background for a reliable estimation of the background level, as the estimated S\'{e}rsic index, $n$, is slightly sensitive to the masked regions. This sensitivity is higher for larger S\'{e}rsic indices. Comparing the estimated effective radius, ($R_e$, i.e. the radius encompassing half-light of the galaxy) and effective surface brightness, ($\langle \mu \rangle_e$, i.e. the mean surface brightness within $R_e$), from Galfit and SExtractor, we found that the difference in results (the scatter around the line with the slope ``one") is minimized when the size of the fitted area (galaxy and background) is about 2.5 times the Kron radius. In each case, the modelled galaxy and the Galfit residual images were inspected by eye to identify galaxies that are well described by S\'{e}rsic model. The galaxies with internal spiral structure or those with poor fit were excluded from the analysis.

To study the dependency of measured S\'ersic indices on the observing wavelengths, surface brightness fitting was performed on both F475W/F814W images. The S\'ersic indices in F814W band are about 10\% higher than those of F475W band (see Figure \ref{fig:ser81475}). To study the wavelength dependency of the FP and PHP in \S \ref{chap:ppanalysis}, for each band, the corresponding S\'ersic index and effective radius were used.

\begin{figure*}
\begin{center}
\includegraphics[width=15cm]{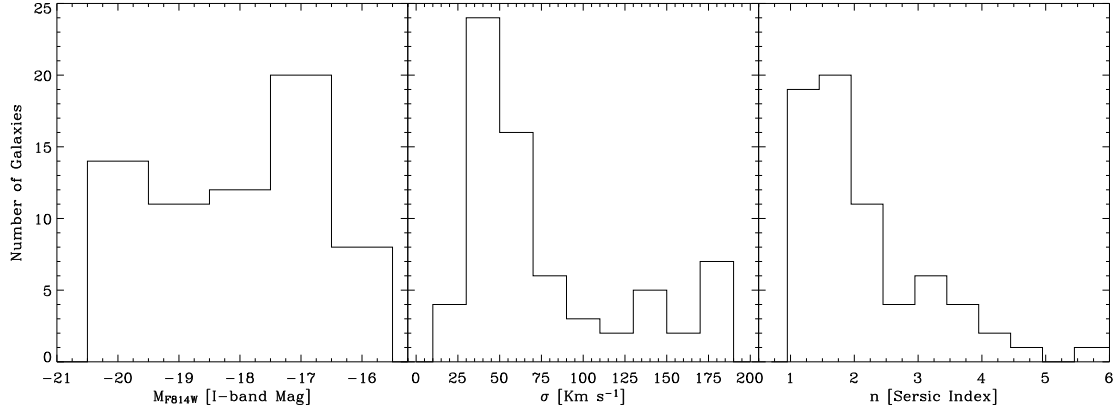}
\caption {
The distribution of galaxies in our sample in terms of the absolute magnitude in F814W-band, central velocity dispersion, $\sigma$, and S\'ersic parameter, $n$.
}
\label{fig:histogram}
\end{center}
\end{figure*}

\begin{figure*}
\begin{center}
\includegraphics[width=13.3cm]{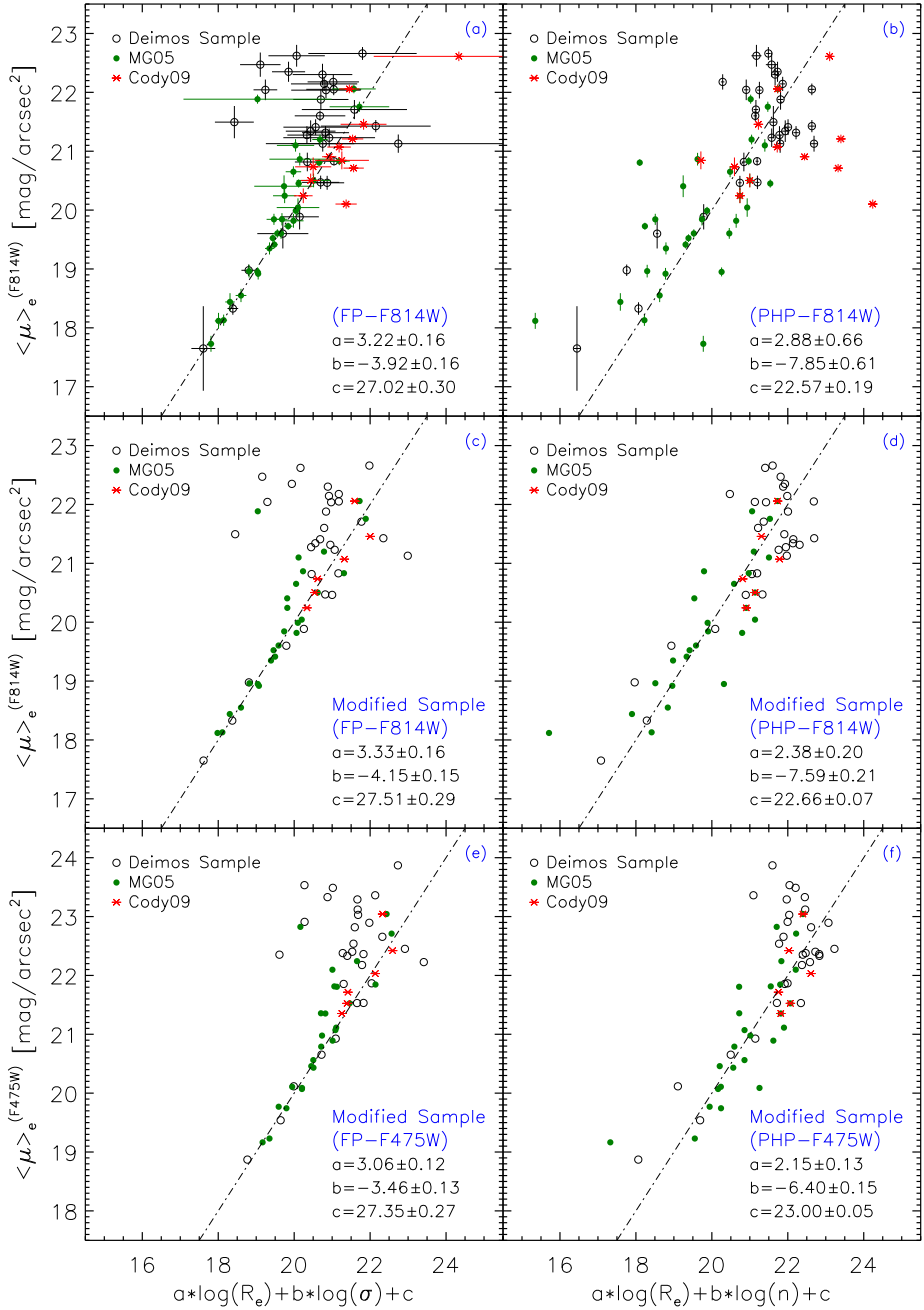}
\caption {
Fundamental and Photometric planes (FP \& PHP) in different filters. {\bf a)} The dash-dotted line represents the projection of the FP fitted for galaxies with $\langle \mu \rangle_e<21$. Most galaxies with $\langle \mu \rangle_e>21$ tend to have higher velocity dispersion compared with that expected from the FP of bright galaxies. $R_e$ and $\sigma$ are in terms of $kpc$ and $km~s^{-1}$, respectively. High errors occur when measuring the velocity dispersion of fainter galaxies (e.g. $\langle \mu \rangle_e>21$) with very low signal-to-noise-ratio in their spectra. Therefore, in this diagram, the horizontal error bars are bigger than the vertical error  bars. {\bf b)} The dash-dotted line represents the projection of the Photometric Plane fitted for galaxies with $\langle \mu \rangle_e<23$. Most of the outliers are S0 galaxies or galaxies with internal structures. These galaxies typically do not have very good S\'{e}rsic fits. All red asterisk outliers belong to dS0 or SB galaxies. {\bf c)} Same as the panel (a) eliminating the galaxies which are not well modelled by S\'ersic function and have internal structures. The modified sample consisting of dwarf ellipticals, dEs, are represented in this panel. To avoid any confusion, the error bars are not shown. The typical size of error bars are the same as panel (a) {\bf d)} Same as the diagram (b) for the same sample as in panel (c). The typical size of error bars are the same as in panel (b). {\bf e and f)} Same as the panels (c) and (d) for F475W-band. All photometric properties and profile parameters (e.g. S\'ersic index) were derived from F475W images. The resulting FP and PHP in both filters, show no discrepancy within 1$\sigma$ error bars. The difference in coefficient ``c'' comes from the different magnitude of each galaxy in different pass-bands. In all panels, open black circles represent dwarf ellipticals from our DEIMOS/Keck observations. For red asterisks and green filled circles, the velocity dispersions were obtained from Cody et al. (2009) and Matkovi\'c \& Guzm\'an (2005), respectively.}
\label{fig:fphotp}
\end{center}
\end{figure*}

\begin{figure*}
\begin{center}
\includegraphics[width=17cm]{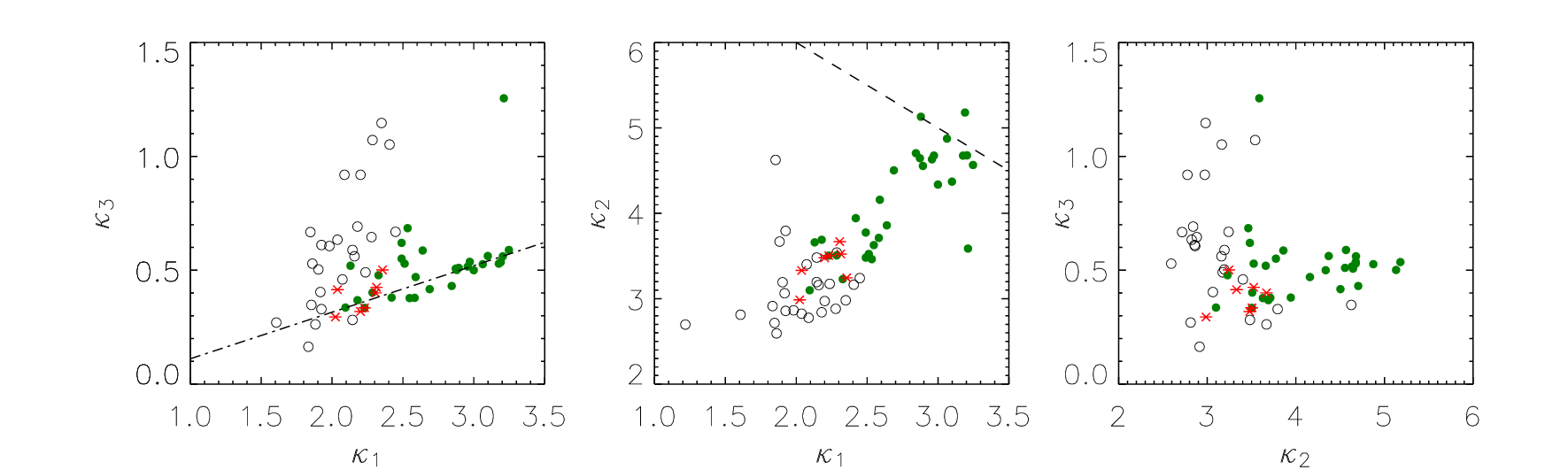}
\caption {
Dwarf elliptical galaxies of the Coma cluster plotted in $\kappa$-space (Bender et al. 1992). All galaxies with bad S\'{e}rsic fit, dS0 galaxies and those with internal spiral structure are not plotted in this diagram. $\kappa_1$, $\kappa_2$ and $\kappa_3$ are proportional to $log(M)$, $log(M/L)\langle \mu \rangle_e^3$ and $log(M/L)$, respectively. The velocity dispersion for black open circles were measured using our DEIMOS data. For red asterisks and filled green circles, the velocity dispersion extracted from Co09  and MG05, respectively. In left panel, the dash-dotted line, ($\kappa_3=(0.20\pm0.06)\kappa_1+(-0.09\pm0.17)$), is the best fit for galaxies with $\langle \mu \rangle_e<20$, for which $\kappa_1$ is typically greater than 2.50. The estimated linear trend for bright galaxies in our sample is consistent with the results of Treu et al. (2006) who found $\kappa_3=(0.21\pm0.02)\kappa_1+(0.19\pm0.08)$ in B-band, for a large sample of Coma galaxies. This diagram indicates that the $M/L$ ratios of less massive dEs are larger than those predicted by the extrapolated line to lower values of $\kappa_1$ ($\kappa_1 \propto log(M)$). The dashed diagonal line in the middle panel ($\kappa_1+\kappa_2=8$) represents the border of the ``zone of avoidance'' not populated by early-type galaxies (Burstein et al. 1997).}
\label{fig:kspace}
\end{center}
\end{figure*}

In addition to Galfit and SExtractor, the IRAF task {\it isophot} in {\it STSDAS} package is used to find and compare the radial light profile of each galaxy and its corresponding estimated S\'{e}rsic model. This helps to examine the reliability of the fitted profiles. Moreover, the central surface brightness of galaxies, $\mu_0$, are obtained from their estimated surface brightness profiles. Figure \ref{fig:galfit} shows the example of two galaxies modelled by S\'{e}rsic function. GMP 3080 seems to have an extra component at its centre which is well modelled by additional Gaussian function with FWHM $\approx 0.13"$, in another iteration. The importance and physical meaning of this excess light is discussed in \S \ref{subsec:extralight}.

The estimated photometric and kinematic parameters of our 71 sample galaxies are presented in Table \ref{tab:results}. Hoyos et al. (2011) present a detailed comparison of effective radius and surface brightness with the ground-based data of Gutierrez et al. (2004) and Aguerri et al. (2005). They find a good agreement, with a few outliers where complex structure is not well resolved in the ground-based data. Our derived values are in general in good agreement with those of Hoyos et al. (2011) although, our derived values of the Se\'rsic index are somewhat lower.  This difference can be attributed to the fact that we allow a separate nuclear component in some of the fits.

\section{The Scaling Relations}
\label{chap:scales}

Taking all essential kinematic and photometric parameters, we have investigated the most well known scaling relations for our sample galaxies. Our sample consists of 71 dwarf galaxies which are fainter and less massive than the previously studied galaxies in the Coma cluster. The distribution of magnitude, velocity dispersion and S\'ersic index of our galaxies are represented in Figure \ref{fig:histogram}. In this section, we also compare the scaling relations of our sample of dEs with other studies.

\subsection{The Fundamental Plane (FP)}
\label{chap:fphotanalysis}

In order to fit the FP to our sample galaxies, we performed an orthogonal distance regression, in which the sum of the orthogonal residuals about the FP is minimized. Compared with the ordinary least-square fitting method, in which the residuals in $\langle \mu \rangle_e$ or $R_e$ are minimized, the orthogonal distance regression is less sensitive to the outliers and is more robust (JFK96; La Barbera el al. 2010). During the fitting process, the square inverse of the error-bars of the measured quantities were used as weight numbers, and therefore the final fitting results are less affected by the values with large error-bars. To do the weighted orthogonal distance regression, we used the {\it ODRPACK} software (Boggs et al. 1989, 1992).

The conventional projection of the FP is shown in Figure \ref{fig:fphotp}. The fitting was performed on bright galaxies (i.e. $\langle \mu \rangle_e^{F814W} < 21$). In panels c and e, we have excluded the galaxies with poor S\'ersic fits. There is no significant difference in the fitted FPs in panels (a) and (c) within 1$\sigma$ uncertainty. 

We probe the dependency of FP on the wavelength by performing the S\'ersic fit to galaxies in F475W filter. In \S\ref{chap:photoanalysis}, we showed that the measured S\'ersic parameter in red filter (F814W) is \%10 higher than the same in blue filter (F475W). Panel (e) in Figure \ref{fig:fphotp} shows the FP for the same galaxies in F475W ACS (g-band) filter. The resulting FP for both red and blue filters (panels c \& e) are compatible with each other within the error bars which implies no colour dependency of the FP of our dEs.

For the 12 brightest galaxies ($M_{814}<-20$ mag) in our sample, the best orthogonal fit for FP is derived as

\begin{align}
\label{fp1}
\nonumber log(R_e)&=(1.33\pm0.02)~log(\sigma)+(0.32\pm0.00) ~ \langle \mu \rangle_e \\
&+(-8.74\pm0.11),
\end{align}

with a RMS scatter of 0.031 dex in log($R_e$). $R_e$ and $\sigma$ are in kpc and km s$^{-1}$, respectively. Even though, our study is based on the precise measurements of the structural and photometric parameters from the HST/ACS imaging data, the coefficients of FP in this analysis for bright galaxies is consistent with those derived from optical (V-band) and near-infrared (K-band) data for 48 giant ellipticals in the Coma cluster (Mobasher et al. 1999: Mo99). Moreover, our resulting FP for bright ellipticals agrees well with other studies of the FP. For instance, JFK96 found the coefficients of $log(\sigma)$ and $\langle \mu \rangle_e$ as 1.3 and 0.32, respectively. In addition, La Barbera et al. (2010) have investigated the FP for a sample of local early-type galaxies ($z<0.1$) from the SDSS DR6, with $M_r \lesssim -20$, in the {\it grizYJHK} wavebands. In agreement with our results, they have shown that the coefficient of $\langle \mu \rangle_e$ is independent of the waveband and equals to 0.32 by minimizing the orthogonal residuals about the FP. The RMS scatter about the FP in previous study of bright ellipticals in the Coma cluster are 0.074 dex in log($R_e$) (Mo99) for optical V-band, 0.08 dex (JFK96; Djorgovski \& Davis 1987) and 0.07 dex (de Carvalho \& Djorgovski 1992) for the B-band and 0.075 dex (Lucey et al. 1991) for the V-band, using other independent samples. The RMS scatter about the fitted FPs depends on the sample completeness and the magnitude range of the studied galaxies. We noticed that adding fainter galaxies results in larger scatters, indicating that faint dwarf galaxies do not lie on the FP of their brighter counterparts.

\begin{center}
\begin{table}

\caption{Coefficients of fundamental plane, FP for different luminosity cutoffs. The FP relation is represented as $log(R_e)=A~log(\sigma)+B~<\mu>_e+C$.}
\centering
\begin{tabular}{| c || c | c | c |}
\hline

 $M_{F814W}$ & A & B & RMS   \\
\hline 
$<-20$ & 1.33$\pm$0.02 & 0.32$\pm0.00$ & 0.031  \\
$<-18$ & 0.77$\pm$0.11 & 0.17$\pm0.03$ & 0.116  \\
$<-16$ & 0.97$\pm$0.16 & 0.19$\pm0.02$ & 0.150 \\
\hline
\end{tabular}
\label{tab:fpstab}

\end{table}
\end{center}

The best coefficients of FP relation for 35 galaxies of our sample, brighter than $M_{814}=-18$, are A=$0.77\pm0.11$ and B=$0.17\pm0.03$ which result in a RMS scatter of 0.116 dex in log($R_e$) corresponding to an uncertainty of 28 per cent in distances to individual galaxies. Considering all galaxies in our sample with $M_{814}<-16$ results in a FP with A=$0.97\pm0.16$ and B=$0.19\pm0.02$ with a RMS scatter of 0.150 dex in log($R_e$), which corresponds to an uncertainty of 35 per cent in distances to individual galaxies. The coefficients of FP for faint magnitude cutoffs, $M_{814}<-18$ and $M_{814}<-16$, differs from those of equation \ref{fp1} for $M_{814}<-20$. This implies that, compared to bright ellipticals with $M_{814}<-20$, the faint dEs lie on an entirely different FP with larger scatter. In addition, We found that the fitted FPs for two sample of faint dEs with different faint-end cutoffs (i.e. $M_{814}<-18$ and $M_{814}<-16$) are consistent with each other within the error bars, suggesting the same FP for dwarf galaxies in the magnitude range $-20<M_{814}<-16$ mag. In Table \ref{tab:fpstab}, the results of the best fitted FPs for different magnitude ranges are presented. We study the departure of faint dEs from the FP of bright ellipticals in \S \ref{chap:departure}.

We also studied the FP in $\kappa$-space (Bender et al. 1992). $\kappa$ parameters form an orthogonal coordinate system which combine the central velocity dispersion, the effective radius and the effective surface brightness. In this formalism, $\kappa1$ and $\kappa3$ are proportional to $log(M)$ and $log(M/L)$ respectively, in which $M$ is the mass of the galaxy while $L$ is its luminosity. Moreover, $\kappa_2$ is proportional to $log(M/L)\langle \mu \rangle_e^3$ and measures the galaxy compactness for a given mass. The projection of the galaxies in $\kappa$-space is illustrated in Figure \ref{fig:kspace}. The left panel of Figure \ref{fig:kspace} shows that the $M/L$ ratio of faint dEs does not decrease on the same linear trend as brighter ellipticals. The variation of the $M/L$ ratio across our sample galaxies and its physical meanings are studied in \S \ref{mtolight}. As seen in the middle panel of Figure\ref{fig:kspace}, the occupied region of our sample dEs in the $\kappa_1-\kappa_2$ diagram agrees well with the Figure 3 of Maraston et al. (2004). This indicates that dEs are generally form a linear trend perpendicular to the occupied region by bulges and giant elliptical galaxies. The different behaviour of giant and dwarf ellipticals in the $\kappa_1-\kappa_2$ space indicates that they are distinct families with different formation scenarios. Similar behaviour, albeit with smaller samples of lower-luminosity galaxies, is seen in Figure 9 of Guzman et al. (1993) and in Figure 1 of Aguerri \& Gonz\'alez-Garc\'ia (2009).

\begin{center}
\begin{table*}
\caption{RMS scatter of the data points about the best fitted fundamental and photometric planes (see Figure \ref{fig:fphotp}). The scatters are calculated along the effective surface brightness, $\langle \mu_{e} \rangle$ and effective radius, $R_e$. For brighter galaxies ($\langle \mu \rangle_e < 21$ in F814W-band) the scatter about the fundamental plane (FP) is smaller than the corresponding photometric plane (PHP). Considering all data points or only fainter galaxies ($\langle \mu \rangle_e > 21$ in F814W-band) indicates that PHP reduces the RMS scatter in both $\langle \mu_{e} \rangle$ and $R_e$.}
\begin{tabular}{|c||c | c||c|c|c||c|c|c|}
\hline

\multirow{2}{*}{Fitted Plane} & \multirow{2}{*}{Filter} & \multirow{2}{*}{Sample} & \multicolumn{3}{c||}{RMS scatter along $\langle \mu \rangle_e$ [mag]} & \multicolumn{3}{c|}{RMS scatter along $log(R_e)$ [dex.]} \\ \cline{4-9}

 &  &  & all data & $\langle \mu \rangle_e < 21$ & $\langle \mu_{e} \rangle > 21$  & all data & $\langle \mu \rangle_e < 21$ & $\langle \mu \rangle_e > 21$ \\ \hline  \hline

\hline
FP (a)	& F814W & all & 1.01 & 0.35 & 1.50 & 0.32 & 0.11 & 0.47 \\
FP (c)	& F814W & modified & 1.04 & 0.28 & 1.49 & 0.31 & 0.08 & 0.45 \\
FP (e)	& F475W & modified &  1.04 & 0.28 & 1.49 & 0.34 & 0.09 & 0.49 \\

\hline \hline
PHP (b)	& F814W & all &  1.10 & 1.24 & 0.87 & 0.38 & 0.43 & 0.30 \\
PHP (d)	& F814W & modified &  0.72 & 0.72 & 0.73 & 0.30 & 0.30 & 0.31 \\
PHP (f)	& F475W & modified &  0.76 & 0.60 & 0.92 & 0.36 & 0.28 & 0.43 \\

\hline
\end{tabular}
\label{tab:deldex}

\end{table*}

\end{center}
\subsection{The Photometric Plane (PHP)}
\label{chap:ppanalysis}

\begin{figure}
\begin{center}
\includegraphics[width=6.5cm]{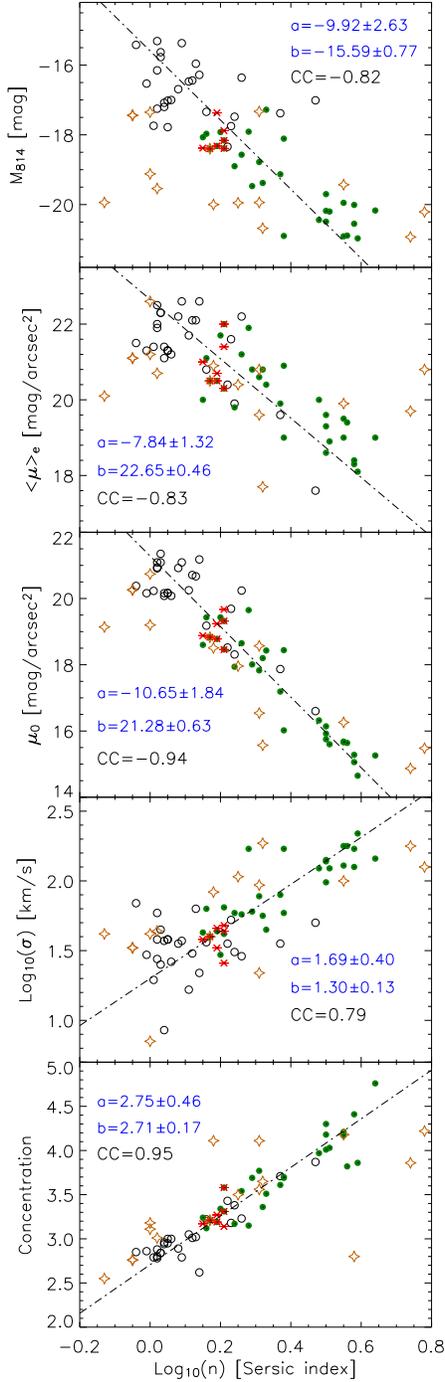}
\caption {
The relation between S\'{e}rsic index, $n$, of the galaxies under study and the photometric and dynamical parameters (i.e. M$_{814}$, $\mu_0$, $\langle \mu\rangle_e$, concentration parameter and $\sigma$). $\langle \mu \rangle_e$ and $\mu_0$ are the effective and central surface brightness of the galaxies, respectively. For definition of concentration parameter refer to \S \ref{chap:photoanalysis}. All photometric parameters are extracted from F814W ACS images. For open circles, the velocity dispersions are derived from DEIMOS spectra in this study and for red asterisks and filled green circles they are extracted from Co09  and MG05 catalogues, respectively. Open stars belong to the galaxies with bad S\'{e}rsic fit or with internal spiral structures. In each panel, `a' and `b' are respectively the slope and intercept of the fitted lines using orthogonal regression. CC is the cross-correlation coefficient of the fitting parameters. }
\label{fig:logn5}
\end{center}
\end{figure}

The S\'ersic index and the central velocity dispersion of galaxies are correlated (see Figure \ref{fig:logn5}) and hence, one is able to use the S\'ersic index instead of the velocity dispersion whenever the kinematic parameters of the galaxies are not available. Replacing the velocity dispersion, $\sigma$, with the S\'ersic index, $n$, in FP relation (Equation \ref{fp}), the photometric plane (PHP) is obtained more economically than the fundamental plane, only based on the photometric properties of the galaxies. Very similarly, PHP could be valuably used as a diagnostic tool to study the galaxy evolution and constrain the early-type galaxy luminosity evolution with redshift (La Barbera 2005: Lb05). In order to provide a local universe reference and to compare the relation between FP and PHP of the early-type galaxies, we construct the PHP for our sample of dwarf galaxies.

Right panels of Figure \ref{fig:fphotp} show the projection of best PHP for our sample galaxies regardless of their brightness. We note that most of the outliers in top right panel have internal spiral structure or poor S\'{e}rsic fit. It appears that while the fundamental plane is less sensitive to detailed morphology of the galaxies, the photometric plane is able to effectively differentiate the morphological type of the galaxies. In right middle panel, we have eliminated the outliers of top panel in order to only have elliptical galaxies with good S\'{e}rsic light profiles (hereafter ``the modified sample''). This reduces the RMS scatter in $\langle \mu \rangle_e$ from 1.12 to 0.74 $mag$. In addition, we have constructed the PHP using F475W ACS (g-band) data in order to examine the wavelength dependency of PHP. The resulting PHP for both filters (panels d \& f) are compatible with each other within the uncertainties. This is also consistent with the results of multi-band (R/I/K-band) study of the PHP of early-type galaxies (Lb05). Considering the following representation for the PHP,

\begin{equation}
log(R_e)=\alpha~log(n)+\beta ~ \langle \mu \rangle_e+\gamma ,
\end{equation}

Lb05 obtained $\alpha \approx 1$ and $\beta \approx 0.2$, both in the optical and near-infrared wavebands, for a sample of galaxies brighter than $M_I\approx-20$ mag located in ``MS1008-1224'' galaxy cluster at $z=0.306$. Lb05 have reported the intrinsic RMS dispersion in $R_e$ as 32\%. Re-deriving the PHP coefficients for the modified GR02 sample of galaxies in Virgo and Fornax clusters and using the same regression method, Lb05 found $\alpha$ and $\beta$ as 1.14$\pm$0.15 and 0.180$\pm$0.031, respectively. This implies no redshift dependence of the PHP of bright elliptical galaxies.

For galaxies brighter than $M_{814}\approx M_I \approx-18$ mag in our modified sample, we found $\alpha=1.82\pm0.10$ and $\beta=0.27\pm0.02$ with a RMS scatter of 0.173 dex in log($R_e$), corresponding to an uncertainty of 40 per cent in distances to individual galaxies. Considering all galaxies brighter than $M_I\approx-16$ the results are $\alpha=2.78\pm0.23$ and $\beta=0.39\pm0.04$, with a RMS scatter of W dex in log($R_e$) or 61 per cent error in distances to each galaxy.

In this study, we have few bright galaxies and therefore we cannot compare our fitted PHP for bright galaxies with results in the literature. The measured coefficients of the PHP relation for our dEs differs from those of previous studies of the PHP of of bright elliptical galaxies (GR02; Lb05). We attribute this discrepancy mainly to the difference in mass range, most of our galaxies have $\sigma < 100 km s^{-1}$ whereas previous samples have $\sigma > 100 km s^{-1}$. The typical S\'ersic indices of our galaxies are less than $\sim$2 while those of GR02 and Lb05 are greater (see Figure \ref{fig:histogram}). As another possibility to explain the discrepancies, we note that our study is based on the HST/ACS photometric data and, therefore the measured photometric and structural parameters are likely to be more accurate than those measured from ground based observations.

\subsection{Fundamental Plane vs. Photometric Plane}
\label{chap:fvpanalysis}

As Figure \ref{fig:fphotp} shows, the deviation of galaxies from the FP and the PHP become larger as their luminosities, central velocity dispersions and S\'{e}rsic indices decrease. In order to quantify the dispersion around the planes, the RMS deviation about the fitted planes are listed in Table \ref{tab:deldex}. The deviations are derived in terms of the mean effective surface brightness and the effective radius for three cases: (i) all data points, (ii) brighter galaxies with $\langle \mu \rangle_e < 21$ and (iii) fainter galaxies with $\langle \mu \rangle_e > 21$. The table shows that the scatter around the PHP is less than the corresponding FP when we consider all of our modified sample galaxies (panels {\it c} and {\it d} of Figure \ref{fig:fphotp}) or when galaxies are fainter than ($\langle \mu \rangle_e > 21$). For our modified sample (i.e. galaxies with good S\'ersic fit or without any internal structure) the RMS scatter in $\langle \mu \rangle_e$ for the FP is $1.04~mag$ while it is reduced to $0.72~mag$ for the PHP. The corresponding deviation in $R_e$ is $0.31~dex$ for the FP and $0.30~dex$ for the PHP. If we only consider the fainter galaxies (i.e. $\langle \mu \rangle_e > 21$), the RMS deviation in $\langle \mu \rangle_e$ is $1.49~mag$ for FP, while PHP decreases this value to $0.73~mag$. The RMS scatter of brighter galaxies (i.e. $\langle \mu \rangle_e < 21$) around the PHP increases compared to the corresponding FP. This implies that the brighter galaxies are still better placed on the FP. We conclude that faint galaxies agree better with the photometric plane of brighter galaxies, compared to the fundamental plane. This implies that the galaxy substructures which are not reflected in velocity dispersion, and hence, in the FP relation, are better reflected in the S\'ersic parameter, and hence, in the PHP relation. In addition, the larger errors in velocity dispersion measurements for fainter dEs may be responsible for their larger scatter about the conventional FP compared to the PHP.

\subsection{Correlations with the S\'ersic index}

In Figure \ref{fig:logn5}, we present the correlations between the S\'{e}rsic index of the sample galaxies and $M_{814}$, $\langle \mu \rangle_e$, $\mu_0$, $\sigma$ and concentration parameter. We find a linear trend between the S\'ersic index and the central surface brightness as $log_{10}(n)=(2.00\pm0.35)-(0.09\pm0.02)\mu_0$ with the correlation coefficient of -W. This is entirely consistent with the relation between these quantities found by Graham \& Guzm\'an (2003). Replacing $n$ by $\sigma$ we find a similar relation (i.e. $log_{10}(\sigma)=(3.97\pm0.56)-(0.12\pm0.02)\mu_0$) with weaker correlation coefficient of -0.80. We ignore the outliers in the fitting process, which are the galaxies with poor S\'ersic fit, represented by open asterisks. Figure \ref{fig:logn5} indicates another correlation between the central velocity dispersion of our sample galaxies, $\sigma$, and their S\'{e}rsic indices, $n$, with the correlation coefficient of 0.79. This correlation enables us to construct a relation $\sigma$ and $n$ as $log_{10}(\sigma)=(0.59\pm0.14)log_{10}(n)-(0.77\pm0.20)$ where $\sigma$ is in km s$^{-1}$.

The correlation between the light concentration in galaxies and their S\'{e}rsic index is also presented in the bottom panel of Figure \ref{fig:logn5}. Ignoring the outliers which are illustrated with open asterisks, this relation is written as $log_{10}(n)=(0.36\pm0.06)C-(0.98\pm0.18)$ with the correlation coefficient of 0.95, where C is the concentration parameter and is defined as $C=5~log_{10}({r_{80}/r_{20})}$. The process of determining the S\'{e}rsic index is model dependent while the concentration parameter is model independent and well determined using a simple photometric analysis. The $C-n$ relation shows that ``$C$" and ``$n$" could be interchangely used.

\begin{figure}
\begin{center}
\includegraphics[width=7.6cm]{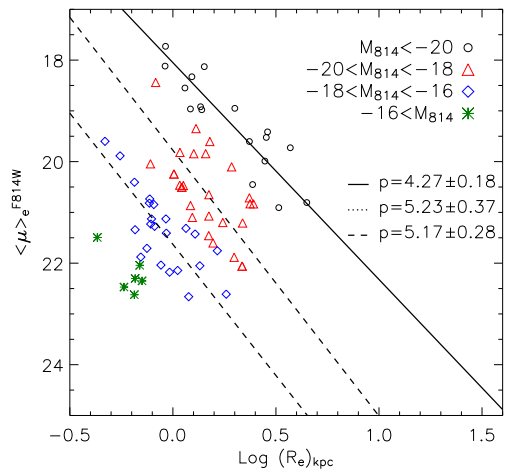}
\caption {
The linear relation between $\langle \mu \rangle_e$ and $log(R_e)$, also known as the Kormendy relation, for different magnitude ranges in F814W-band. The coefficient ``p'' is the slope of the linear trends: $\langle \mu \rangle_e=p ~ log(R_e)+q$.}
\label{fig:kormendy}
\end{center}
\end{figure}

\subsection{The size-surface brightness relation}
The linear relation between the effective surface brightness, $\langle \mu \rangle_e$, and half-light radius, $R_e$, of elliptical galaxies, also known as Kormendy relation (KR, Kormendy 1977), is represented as 

\begin{equation}
\langle \mu \rangle_e=p ~ log(R_e)+q.
\end{equation}

Figure \ref{fig:kormendy} shows the best fit of KR for our dEs in different luminosity ranges, $M_{814}<-20$, $-20<M_{814}<-18$ and $-18<M_{814}<-16$. For our brightest galaxies ($M_{814}<-20$), we find the KR slope as 4.27$\pm$0.18 which is significantly comparable to 2.43$\pm$0.15 for a sample of giant galaxies in Coma cluster with central velocity dispersion $\sigma>200$ km s$^{-1}$ (Ziegler et al. 1999). We attribute this discrepancy to the fact that our sample covers different range of size and magnitude compared with the galaxies studied by Ziegler et al. (1999). The typical velocity dispersion of our dEs is less than 100 km s$^{-1}$ and the average size of our dEs is $\sim$1.5 Kpc. Moreover, the effective radii of our galaxies with $M_{814}<-20$ are less than $\sim$4.5 Kpc, while all other studies of KR cover galaxies with larger sizes ($>\sim 10 $ Kpc) in this magnitude range (e.g. La Berbera et al. (2010); Ziegler et al. 1999; D'Onofrio et al. 2006). The imposed systematic restrictions on our sample of dEs, such as luminosity and size cuts, changes the geometric shape of the distribution of galaxies on the log($R_e$)-$\langle \mu \rangle_e$ plane, and therefore results in different KR slope (Nigoche-Netro et al. 2008). Fitting the KR to our dEs in the magnitude range $-20<M_{814}<-18$ and $-18<M_{814}<-16$, we found the KR slope as 5.23$\pm$0.37 and 5.17$\pm$0.28, respectively. In agreement with our results, Khosroshahi et al. (2004) have also found the slope of KR as $\sim$5.2 for dwarf ellipticals of 16 nearby galaxy groups with $-14<M_R<-18$.

The KR is originated from the definition of the effective radius, $R_e$, which relates the luminosity and effective surface brightness as $L=2\pi I_eR_e^2$. Theoretically, $p$ equals 5, and any difference from 5 is the results of the change in geometric shape of the distribution of galaxies on the log($R_e$)-$\langle \mu \rangle_e$ plane. Any change in magnitude range of the galaxies and the shape of the magnitude distribution results in different slope of KR (Nigoche-Netro et al. 2008). 

In agreement with Khosroshahi et al. 2004 and D'Onofrio et al. (2006), Figure \ref{fig:kormendy} shows that for fainter galaxies the log($R_e$)-$\langle \mu \rangle_e$ linear trend is steeper than that of the brighter galaxies ($M_{814}<-20$). We also noticed that the slopes of KR for our galaxies in the magnitude range $-20<M_{814}<-18$ and $-18<M_{814}<-16$ are consistent within the error bars. Due to the limited $R_e$ range, we did not fit the KR for galaxies with $-16<M_{814}$.

\subsection{The size-luminosity relation}

\begin{figure}
\begin{center}
\includegraphics[width=8.5cm]{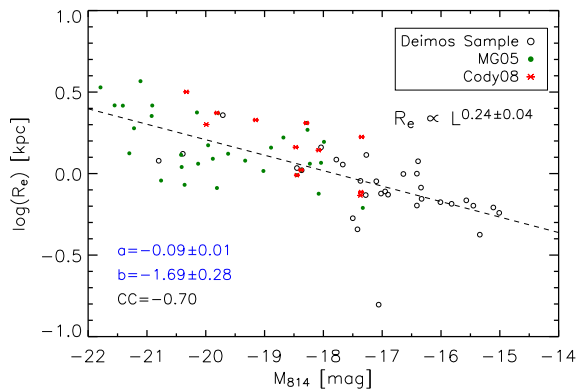}
\caption {The effective radius against the luminosity of our sample galaxies in F814W-band. `a' and `b' are slope and intercept of the best linear trend which is found using an orthogonal distance regression (dashed line). As seen, $R_e$ and luminosity are correlated with the correlation coefficient of CC=-0.70. The RMS scatter about the fitted line is 0.17 dex. along the vertical axis, $R_e$.
}
\label{fig:rem}
\end{center}
\end{figure}

In Figure \ref{fig:rem}, we show the size-luminosity relation, $R_e \propto L^\tau$, for our sample dEs. We found $\tau \approx 0.24$, which is consistent with the results of de Rijcke et al. (2005) who found the B-band radius-luminosity power-law slope between 0.28 and 0.55 for a sample of dwarf ellipticals and dwarf spheroidal galaxies. $\tau$ depends on the luminosity range of the galaxies under study. As Desroches et al. (2007) found, $\tau$ systematically varies from $\tau \approx 0.5$ at $M_r \approx -20$ to $\tau \approx 0.7$ at $M_r \approx -24$, which are in general in agreement with our result if one extrapolates them to lower luminosities despite the structural differences between luminous and faint dEs.

\begin{figure*}
\begin{center}
\includegraphics[width=13cm]{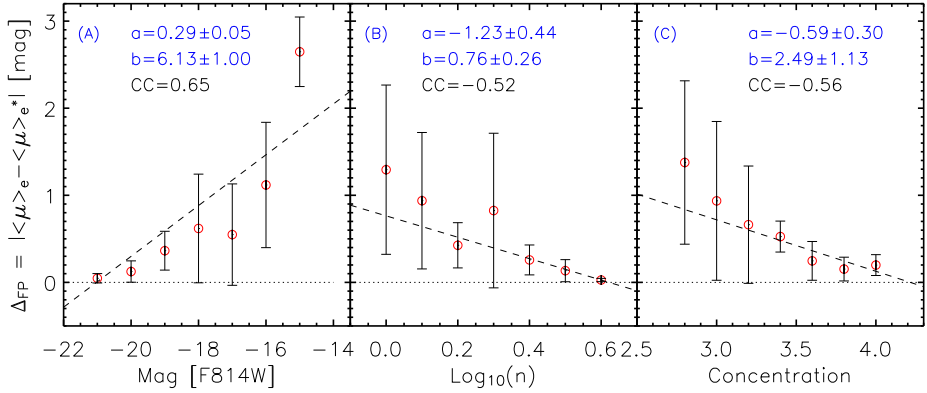}
\caption {
The deviation of the real effective surface brightness, $\langle \mu \rangle_e$, from that predicted by FP, $\langle \mu \rangle_e^*$, against different dynamical and photometric  parameters. Panel (A) shows the tight correlation between the deviation from the FP and the luminosity of the galaxies. Panel (B) represents weaker correlations in terms of $n$. In panel (C), the deviation is drawn in terms of the concentration parameter which is model independent and defined as $5*log_{10}(r_{80}/r_{20})$. For all cases we used the least square method to find the best linear trend (dashed lines) by minimizing the residuals in $\Delta_{FP}$. `a' and `b' are slope and intercept of the fitted lines, respectively.
}
\label{fig:deviation0}
\end{center}
\end{figure*}

\section{Departure of Galaxies from the FP }
\label{chap:departure}

In order to understand the deviation of the faint dEs from the FP associated with the brighter galaxies, we study the correlations between the departure of galaxies from the FP with other observables (e.g magnitude, S\'ersic index, concentration parameter, velocity dispersion, colour). In this study, we define the departure of galaxies from the FP as $\Delta_{FP}=|\langle \mu \rangle_e-\langle \mu \rangle_e^*|$, where $\langle \mu \rangle_e$ is the effective surface brightness of the sample galaxies and $\langle \mu \rangle_e^*$ is the expected effective surface brightness from the FP of bright ellipticals.

\subsection{Dependence on Galaxy Light Profile}

In Figure \ref{fig:deviation0}, the correlation between $\Delta_{FP}$ and magnitude, $M_{814}$, S\'ersic index ($n$), and the concentration parameter are studied. The data points are binned and the error bars are based on the 1$\sigma$ RMS scatter of $\Delta_{FP}$ within each bin.  As different panels of this Figure show, $\Delta_{FP}$ is anti-correlated with the luminosity of galaxies, their S\'ersic indices and their light concentration. Although these three parameters are correlated with each other (see Figure \ref{fig:logn5}), this graph shows that the maximum correlation is seen between the luminosity of galaxies and $\Delta_{FP}$ with a correlation coefficient of 0.65.

In this study, we obtain the partial correlation coefficients between $\Delta_{FP}$ and other observables, defined as $P_{cc}(\Delta_{FP},\alpha; \beta)$. Here, $P_{cc}$ measures the correlation between $\Delta_{FP}$ and one of the observables, $\alpha$, in the case where the influence of the third variable, $\beta$, is eliminated. We obtained the partial correlation coefficients between $\Delta_{FP}$ and luminosity by taking out the effects of $\sigma$ and $n$ as $P_{cc}(\Delta_{FP},M_{814};\sigma)$=0.64 and $P_{cc}(\Delta_{FP},M_{814};n)$=0.45. In addition, we found that $\Delta_{FP}$ is very weakly correlated with $\sigma$ with the correlation coefficient of -0.30. This is consistent with the fact that $\sigma$ itself contributes in the FP relation and hence no strong correlation between $\Delta_{FP}$ and $\sigma$ is expected. As another fact, taking out the effect of $\sigma$ (or $n$), when calculating the $P_{cc}(\Delta_{FP},M_{814};\sigma ~or~ n)$, does not significantly change the correlation coefficient between $\Delta_{FP}$ and $M_{814}$. Furthermore, the correlation coefficient between $\Delta_{FP}$ and $n$ is -0.52 and after taking out the effect of galaxy magnitudes, we obtain $P_{cc}(\Delta_{FP},n;M_{814})$=-0.19. We also obtained $P_{cc}(\Delta_{FP},n;\sigma)$=-0.46. These all imply that the departure from the FP is correlated more tightly with luminosity than with the S\'{e}rsic index.

The departure from the FP, $\Delta_{FP}$, is also correlated with the light concentration in the sample galaxies, C, the cross correlation factor of -0.56. As the concentration is expected to be correlated with the intrinsic luminosity of the galaxies, we found partial correlation to be $P_{cc}(\Delta_{FP},C;M_{814})$=-0.17. This is similar to the $\Delta_{FP}$ correlation with the S\'{e}rsic index, as both indicate roughly the same in galaxy light profile. This can also be inferred from the tight correlation between S\'ersic index and concentration parameter (see Figure \ref{fig:logn5}).

We have also examined the correlation between $\Delta_{FP}$ and effective radius, $R_e$, of our sample galaxies, as a representative of their size. In this case, the correlation coefficient is obtained as -0.30. This implies that $\Delta_{FP}$ is almost size independent. This is also expected form the contribution of $R_e$ in the FP relation.

\subsection{Dependence on Galaxy Central Excess Light}
\label{subsec:extralight}

Another possible parameter which may be correlated with $\Delta_{FP}$ is the central excess light relative to their best single S\'ersic model (see Figure \ref{fig:galfit}). This central excess light (CEL), or nuclear star cluster, occurs in the majority of dwarf galaxies in the Coma cluster (den Brok et al. 2011), and is somewhat bluer than the main body of the galaxy. The excess light is an imprint of the formation history of galaxies and its colour and distribution indicate that it is a result of some dissipational process. A probable explanation is that it is the result of the last major merger, in which the gas content of the gas-rich progenitor falls into the core and undergoes star formation generating the excess light (e.g. HCH08, Kormendy et al. 2009). Numerical simulations (Makino \& Hut 1997)  show that the galaxy merger rate scales as $N^2$/$\sigma^3$, where $N$ is the density of galaxies, and $\sigma$ is velocity dispersion. Hence in Coma, with 3.5 times the  density and $\sigma$ 1.2 times greater, mergers should occur 7 times more frequently than in Virgo. den Brok et al (2011) conclude that the observed relationship between S\'ersic index and the strength of the colour gradient also points towards a history of wet mergers in the history of dwarf galaxies.

Alternatively the nuclear excess could be formed from gas already in the dwarf galaxies as they fall into the cluster. Ram pressure stripping will remove gas from the outer parts and tidal interactions could drive the residual gas into the core, where it forms stars. In this case we would expect an environmental dependence of the CEL properties. Such dependence will be complex, as both the stripping which removes gas, and the tidal interactions driving residual gas to the core, will be strongest in the centre of the cluster, unfortunately due to the failure of ACS in 2007, and the consequent truncation of the HST survey, the sample of dwarfs with high spatial resolution images outside the cluster core is small.

An inspection the radial surface brightness profile of our sample galaxies confirmed that most part of the galaxies, beyond the aperture with $0.40$ arcsec diameter, are well modelled by S\'ersic function. Therefore, we measured the central excess light of our sample galaxies within apertures with diameter of 0.4 arcsec (i.e. ~0.19 kpc at Coma distance), which were co-centred with the target galaxies. Here, we represent the central excess light as

\begin{figure}
\begin{center}
\includegraphics[width=7.5cm]{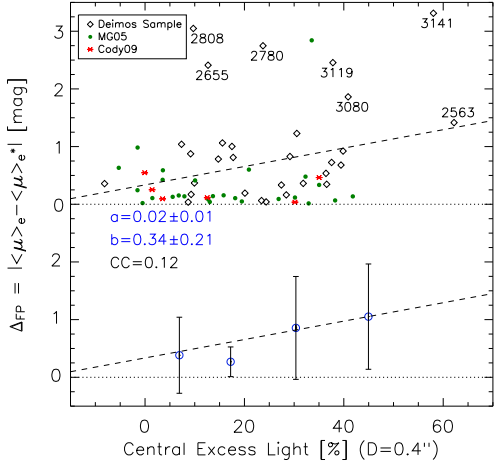}
\caption {
Deviation from the FP against the relative central excess light (i.e. CEL, see the text). CEL is measured within an aperture of $0.4"$ diameter (i.e. $\sim$0.19 kpc). The black open diamonds show the galaxies in our DEIMOS sample. For red asterisks and filled green circles, the velocity dispersion were extracted from Co09 and MG05, respectively. The most deviant galaxies are labelled with their GMP IDs (1983).}
\label{fig:excess}
\end{center}
\end{figure}

\begin{align}
\label{extra}
CEL=\left(\frac{F_{gal}-F_{mod}}{F_{mod}}\right)_{aper},
\end{align}

where, $F_{gal}$ is the flux of the target galaxy within the defined central aperture, and $F_{mod}$ is the flux of the fitted S\'ersic model, within the same aperture. The extra flux is plotted against the deviation of galaxies from the FP ($\Delta_{FP}$) in Figure \ref{fig:excess}. Even though the $\Delta_{FP}-CEL$ correlation coefficient is 0.12 and we did not find any clear relation between $\Delta_{FP}$ and CEL. In general, the most deviant galaxies (the labelled points) tend to have more central excess light. As Figure \ref{fig:excess} shows, the relative CEL of all galaxies with $\Delta_{FP}>1$ mag are more than 10\%.

It is likely that these galaxies have extra luminosity at their central regions due to star formation imposed by either wet mergers of gas driven to their centres by tidal interactions. The star formation activity expels out the luminous baryonic matter from the galaxy and changes its mass-to-light ratio, $M/L$. Any change in the $M/L$ ratio results in the deviation of the galaxy from FP. We study the $M/L$ ratio of our sample dEs in \S \ref{mtolight}.

\begin{figure*}
\begin{center}
\includegraphics[width=16cm]{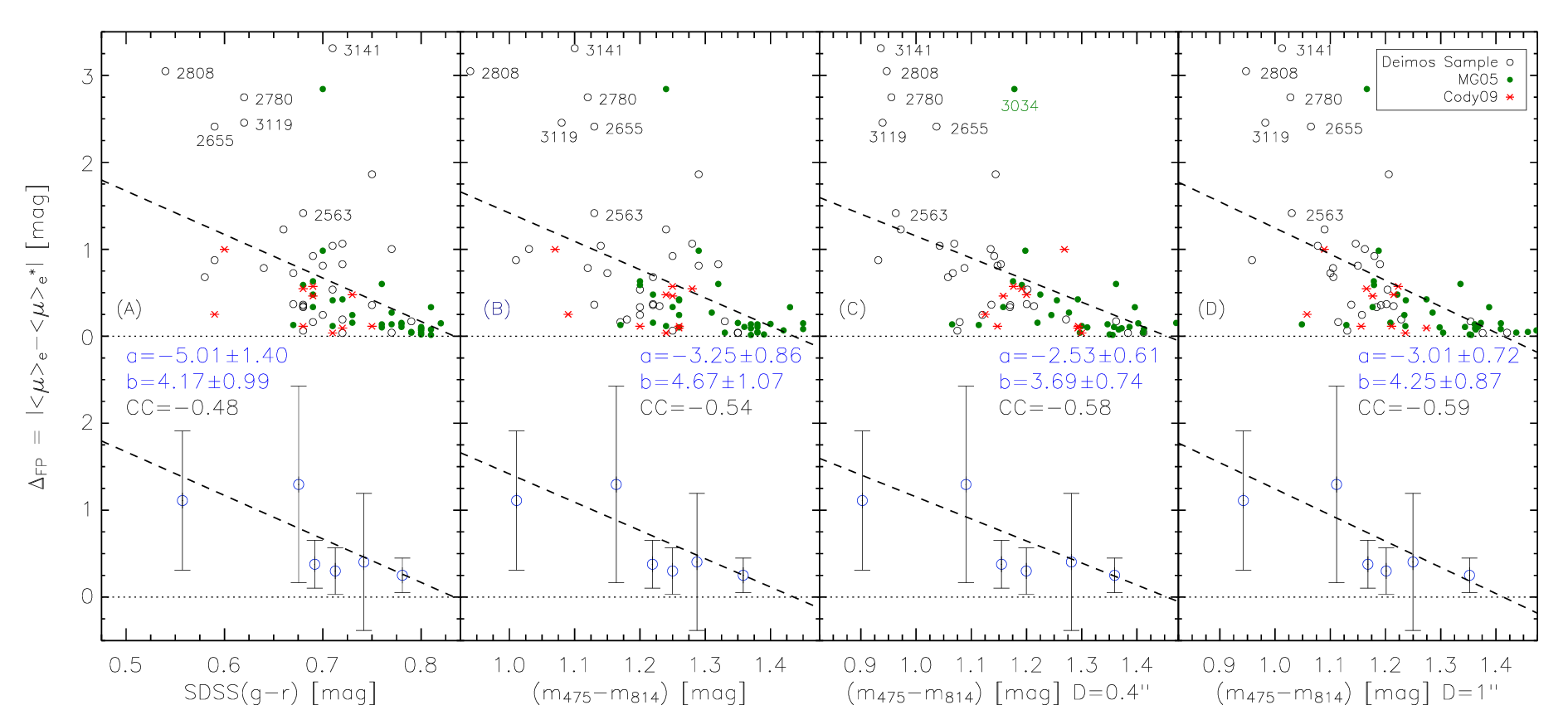}
\caption { Deviation of galaxies from the FP (F814W-band), $\Delta_{FP}$, against their colours. The colour of galaxies in panel (A) were calculated based on SDSS DR7. In panel (B), colours were derived from our analysis on ACS F814W/F475W images. Panel C \& D are the same as panel (B) except for colours which were derived within the apertures of 0.4'' and 1'' diameter at the centre of galaxies. At the bottom of each panel, all data points were binned. Each bin contains the same number of galaxies. For all cases we used the least square method to find the best linear trend (dashed lines) by minimizing the residuals in $\Delta_{FP}$. `a' is the slope and `b' is the intercept of the fitted lines. CC is the cross-correlation coefficient of the fitting parameters. This diagram shows that the bluer galaxies are more deviant from the FP specifically when considering the colour of galaxies at their 0.4'' or 1.0'' central regions. The most deviant galaxies are labelled with their GMP IDs.}
\label{fig:deviation}
\end{center}
\end{figure*}

\subsection{Dependence on Galaxy Colour}
\label{sec:devcol}

The location of early-type galaxies relative to the FP is a result of their age and stellar population properties (Graves et al. 2010; Graves et al. 2009, Terlevich \& Forbes 2002; Forbes at al. 1998). Here, we study the colour-$\Delta_{FP}$ relation, in order to assess the stellar population dependency of the faint dEs deviation from the FP of bright ellipticals.

The colour of our dEs are calculated in 4 forms. First, we used the SDSS DR7 data to construct $g-r$ for all of our galaxies. Second, we used our HST/ACS images in two F814W (I-band) and F475W (g-bang) to calculate $\Delta m = m_{475}-m_{814}$. In addition, we estimated the galaxy colours within two co-centred apertures with diameters of 0.4 and 1.0 arcsec (equal to 0.30 and 0.75 kpc), located at the centre of each galaxy (hereafter, $\Delta m(04)$ and $\Delta m(10)$, respectively). In table \ref{tab:crosscor}, we present the correlation factor for each pair of defined colours of our sample galaxies and other photometric and dynamical  parameters (i.e. $M_{814}$, $\sigma$, n and $\Delta_{FP}$).

\begin{table}
\caption{The cross correlation coefficient of the colour of our sample galaxies and other photometric and dynamical parameters. Please refer to \S \ref{sec:devcol} for definition of colour parameters. Figure \ref{fig:deviation} also represents $\Delta_{FP}$ versus the Colour parameters.   }
\begin{center}
\begin{tabular}{c | c c c c c}\hline
{\it Correlation} & $\Delta_{FP}$ & $M_{814}$ & $log_{10}(\sigma)$ & $log_{10}(n)$ \\
{\it Coefficients} & (1) & (2) & (3) & (4) \\ 
\hline
g-r & -0.48 & -0.73 & 0.57 & 0.72 \\
$\Delta m$ & -0.54 & -0.78 & 0.58 & 0.74 \\
$\Delta m(04)$ &  -0.58 &  -0.78 & 0.61 & 0.74 \\
$\Delta m(10)$ & -0.59 & -0.81 & 0.62  & 0.78 \\
\hline
\end{tabular}
\end{center}
\label{tab:crosscor}
\end{table}

Table \ref{tab:crosscor} shows that $\Delta_{FP}$ is tighter correlated with the colour of galaxies compared to their central excess light (see \S \ref{subsec:extralight}). In all cases, $\Delta m(10)$ displays stronger correlation compared to all other defined colour parameters. Moreover, $\Delta m(10)$ is stronger correlated with the galaxy magnitudes, $M_{814}$, implying that fainter galaxies are bluer. A strong correlation also exists between $\Delta m(10)$ and $log_{10}(n)$, which indicates that bluer galaxies have smaller S\'ersic indices and therefore are less concentrated.

In Figure \ref{fig:deviation}, the deviation of galaxies from the FP is plotted against different colour indicators. 
As seen, the bluer galaxies have larger $\Delta_{FP}$ values. In agreement with our results, Bernardi et al. (2003) have also found a correlation between the the residuals to the FP and colour of their sample of bright early-type galaxies. Figure 3 of their paper indicates that bluer galaxies display greater $\Delta_{FP}$.

This implies that the deviation of dEs from the FP is correlated to their stellar population. The supernovae activities, as a consequence of star formation in bluer galaxies, has swept away the luminous baryonic matter from these galaxies. Thus the supernova driven winds truncate star formation and modify the $M/L$ ratio of the bluer galaxies, increasing their scatter about the FP. This scenario also explains the location of the faintest galaxies of our sample on the L-$\sigma$ diagram (see Paper I), indicating that these galaxies have larger velocity dispersions compared to the trend of brighter dEs. We noted that, these galaxies are bluer than the other galaxies in our sample and display larger deviation from the FP. The star formation have ejected the luminous matter of these galaxies and shifted them to the faint end of the L-$\sigma$ diagram, while their total mass (mostly the dark matter content), and hence, their internal velocity dispersions remain the same.

\section{The $M/L$ ratios }
\label{mtolight}

The behaviour of $M/L$ ratio may account for FP tilt (e.g. ZGZ06; Cappellari et al. 2006; Grave \& Faber 2010) and/or deviation of galaxies from the FP (e.g. Reda, Forbes \& Hau 2005). Based on the virial theorem, the dynamical mass, M, within the effective radius, $R_e$ can be derived as,

\begin{align}
\label{massdynamo}
M\equiv c\frac{\sigma^2 R_e}{G}, 
\end{align}

where $\sigma$ is the galaxy central velocity dispersion and G is the gravitational constant. Assuming that our sample is homologous, $c$ is almost constant for our galaxies and depends on the profiles of the dark and luminous matters. We adopt $c\equiv5$ (Cappelari et al. 2006); although our conclusion is independent of the exact value of $c$ as long as it is uniformly applied to all galaxies. In this study, we assumed that the absolute magnitudes of the Sun (in AB system) in F475W and F814W bands are 5.14 and 4.57 mag, respectively.

\begin{figure}
\begin{center}
\includegraphics[width=8cm]{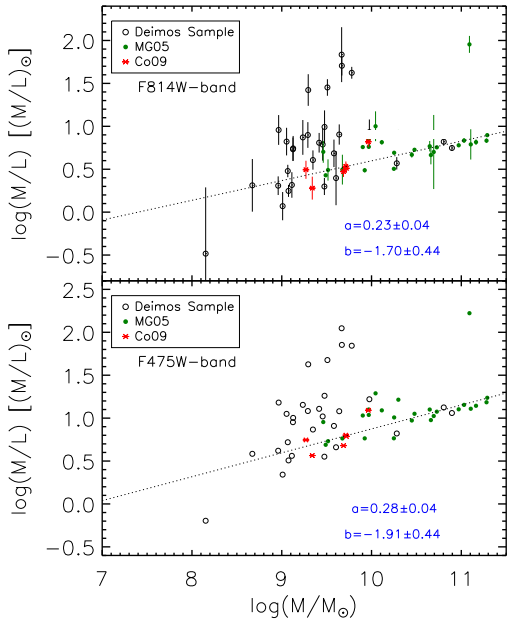}
\caption{The mass-to-light ratio ($M/L$) of our sample galaxies in terms of their dynamical mass. Panels (A) and (B) are based on ACS F814W-band and F475W-band photometric data, respectively. As seen, the slopes of ($M/L$)-$M$ lines do not depend on the two passbands (F814W and F475W). Open circles represent the galaxies with DEIMOS data for which their $\sigma$ were measured in paper I. Red asterisk and green filled circles belong to dEs from Co09 and MG05, respectively. The dotted line is the best linear fit for brighter galaxies with $M_{814}>-17.5$. `a' and `b' are the slope and the intercept of the dotted lines, respectively. In panel (A), the typical error bars of $M/L$ are illustrated. The mass, $M$, and $M/L$ ratios are in solar units.}
\label{fig:mtolratio}
\end{center}
\end{figure}

In Figure \ref{fig:mtolratio}, we show the variation of the dynamical mass-to-light ratio, $M/L$, as function of galaxy dynamical mass, M. For galaxies brighter than $M_{814}=-17.5$ mag, the correlation between $M/L$ and $M$ (the dotted line) is obtained as $M/L \propto M^{0.23\pm0.04}$, in agreement with the results of other studies for bright ellipticals. As an example, Mo99 found $M/L \propto M^{0.18\pm0.01}$ in K-band and $M/L \propto M^{0.23\pm0.01}$ in optical V-band, for galaxies in the mass range of $10^{10}$-$10^{13}M_{\odot}$. 
In addition, studying a sample of 9000 local early-type galaxies of SDSS in the redshift range 0.01$<z<$0.3, Bernardi et al. (2003) found $M/L \propto M^{\sim 0.2}$.

In our sample of dEs, the best linear relation between $log_{10}(\sigma)$ and $\langle \mu \rangle_e$ is found to be $log_{10}(\sigma)=(4.94\pm0.75)-(0.16\pm0.02)\langle \mu \rangle_e$ with the correlation coefficient of -0.66. This relation is translated to $I_e \propto \sigma^{2.5\pm 0.3}$, where $I_e$ is effective surface brightness in flux units, in F814W-band. Using the results of Binggeli, Sandage \&  Tarenghi (1984), we expect $I_e \propto \sigma^{-2.5}$ for elliptical galaxies. Using the virial theorem for spherical systems, $\sigma^2 \propto GM_e/R_e$, and the relation between luminosity and effective surface brightness, $L=2\pi I_eR_e^2$, the $M/L$ ratio is obtained as $M/L \propto {\sigma^2}/{\sqrt{I_eL}}$ (Co09). Substituting our derived Faber-Jackson relation (Paper I), $L\propto{1.99\pm0.14}$, and the dependency of $I_e$ upon $\sigma$, we obtain $M/L \propto \sigma^{-0.25\pm0.33} \propto L^{-0.13\pm0.17}\propto M^{-0.15\pm0.22}$ for all galaxies in our sample. Our result is more consistent with Co09 who found the $M_{dyn}/L \propto M^{0.09\pm0.06}$ for a sample with almost the same luminosity range as ours.

As seen in panels (A) \& (B) of Figure \ref{fig:mtolratio}, for $M_{814}>-17.5$, some galaxies have larger $M/L$ ratios with respect to the faintward extrapolation of the linear trend of brighter galaxies, while few faint galaxies follow the $M/L-M$ relation of bright galaxies. The most deviant galaxies in $M/L-M$ diagram (see Figure \ref{fig:mtolratio}) are the bluer galaxies, too. This suggests that the formation of the most deviant dwarf galaxies and those which are following the trend of bright ellipticals can be explained by different scenarios. Nevertheless, we still need more data point to study the scatter of faint dEs about the trends of brighter ellipticals and to examine different formation mechanisms at the faint regime. In agreement with our results (see Figure \ref{fig:mtolratio}), studying the dwarf galaxies with $-16<M_V<-12$ in the core of Perseus cluster, Penny et al. (2009) found that fainter dwarfs have larger $M/L$ ratios. In addition, Geha et al. (2002) showed that fainter galaxies have larger $M/L$ ratios compared to their brighter counterparts

We have also investigated the waveband dependency of ($M/L$)-$M$ relation, to explore whether it is governed by the change in metalicity of galaxies and/or their stellar population. Comparing panels (A) and (B), which are based on ACS F814W and F475W-band photometry, we find the same behaviour of $M/L$ ratio in different wavebands. We noted that, the ($M/L$)-$M$ relation of galaxies fainter than $M_{814}=-17.5$ is also independent of the passband. Therefore, the stellar population is not the only parameter responsible for the change of $M/L$ ratios (Mo99).

As a conclusion, we found that the $M/L$ ratio is not constant over all of our sample dEs and varies with the mass and the luminosity of the galaxies. The variation in $M/L$ ratio is responsible for the deviation of our fainter dEs from the FP. Since the fainter galaxies in our sample are bluer than the other galaxies and have larger $M/L$ ratios, we attribute their deviation from the FP to their recent star formation activities (see also \S \ref{sec:devcol}).

\section{Discussion}
\label{discussion}

In this paper, we studied the fundamental and photometric planes of a sample of 71 dEs in the core of Coma cluster, the nearest massive elliptical-rich cluster down to luminosity of $M_{814}<-15.3$. Taking advantage of the DEIMOS high resolution spectrograph, which enables us to measure the internal velocity dispersion of dwarf ellipticals, and high resolution imaging of HST/ACS which allows an accurate surface brightness modelling, we were able to extend the FP of galaxies to $\sim$1 magnitude fainter than the previous studies.

We obtained the FP for a subsample of 12 galaxies brighter than $M_{814}=-20$ as $R_e \propto \sigma^{1.33\pm0.02} \langle I\rangle_e^{-0.80\pm0.01} $ which is consistent with the previous studies of bright galaxies in Coma (JFK96; Mo99). Studying the FP of 141 early-type galaxies in the Shapley super cluster at $z$=0.049, Gargiulo et al. (2009: Ga09) found that the FP follows the relation $R_e \propto \sigma^{1.35} \langle I\rangle_e^{-0.81}$ for galaxies with $\sigma>100~km~s^{-1}$ and $M_R<-18.7$. When including all galaxies in their sample, including low-mass galaxies down to $\sigma \sim$50 $km~s^{-1}$, Ga09 found a shallower exponent for $\sigma$. The FP relation of our dEs displays even shallower exponents for $\sigma$ and $\langle I \rangle_e$ than in Ga09, due to further extension to fainter galaxies.

In Figures \ref{fig:fphotp} and \ref{fig:deviation}, the faintest galaxies and the most deviant data points are based on our DEIMOS observations. One could argue that our $\sigma$ measurements might be overestimated. In paper I, we have considered all source of uncertainties in measuring the velocity dispersions, namely the statistical errors, template mismatch and other systematic uncertainties. We have also checked the sensitivity of our measured velocity dispersions to different kind of stellar templates. Moreover, using different set of mixed stellar templates, covering a vast range of spectral types, when measuring the velocity dispersions, confirmed our measurements. The presented uncertainty in the measurement includes all such sources and therefore we can rule out the overestimation argument.

In this study, the spectroscopy was carried out using the slits with the size of 0.7 arcsec covering the body and the central region of the galaxies. Due to the size of slits, for galaxies with net rotation, the measured velocity dispersion differs from the central velocity dispersion. The kinetic energy per unit mass (KE) for a spheroid is (e.g. Busarello, Lango \& Feoli 1992):

\begin{align}
\label{ke}
KE=\frac{1}{2}\langle v \rangle = \frac{1}{2}V_{rot}^2+ \frac{3}{2} \sigma_0^2 \simeq \frac{3}{2} \sigma_m^2 ,
\end{align}

where $V_{rot}$ is the rotation velocity, $\sigma_0$ is the central velocity dispersion and $\sigma_m$ is the measured velocity dispersion. Thus we have

\begin{align}
\label{sigma}
(\frac{\sigma_m}{\sigma_0})^2 \simeq 1+\frac{1}{3}(\frac{V_{rot}}{\sigma_0})^2 \ .
\end{align}

The observed values of $V_{rot}/\sigma_0$ range from 0.01 to $\sim$1.0 for galaxies with $-14<M_R<-26$ (Davies et al. 1983; Pedraz et al. 2002; Simien \& Prugniel 2002; Geha et al. 2003; van Zee et al. 2004), implying that the difference between $\sigma_m$ and $\sigma_0$ is less than $\sim$15\% of the central velocity dispersion. This kind of uncertainty is smaller than the total estimated uncertainty in measuring the velocity dispersions of faint dEs ($\sigma \lesssim 50$ km s$^{-1}$). Based on the relation for an oblate, isotropic galaxy flattened by rotation (Davies et al. 1983), the mean isophotal ellipticity of four galaxies in our sample, GMP2780, GMP2563, GMP3119 and GMP3141 are less than 0.1, corresponding to $V_{rot}/\sigma_0$ $\simeq$ 0.3. Using equation \ref{sigma}, the uncertainty in the measured velocity dispersion due to the rotation would be $\sim$2\%, which is significantly smaller than the uncertainties due to template mismatch and other systematics discussed in paper I. The uncertainty due to the rotation is $\sim$6\% for GMP2655 and GMP2808, for which the mean ellipticity is $\sim$0.25. Thus, rotation can not be solely responsible for the deviation of faint dEs from the faintward extrapolation of the Faber-Jackson relation and the FP of bright ellipticals.

The non-universality of the FP is claimed by ZGZ06, where they suggest that the spheroids ranging from dEs to the galaxy clusters lie on a curved surface, in ($\sigma$,$R_e$,$\langle \mu \rangle_e$) space. Describing the $M/L$ ratio as a parabolic function of the internal velocity dispersion, ZGZ06 could explain the change in the coefficients of the FP for different spherical systems. They predicted that for $\sigma \sim 50~ km~s^{-1}$, $M/L$ is almost a flat function of $\sigma$, with a large scatter, which tends to increase for smaller values of $\sigma$. In agreement with Co09 and ZGZ06, we found $M/L \propto \sigma^{-0.25\pm0.33}$ for all galaxies in our sample. In addition, we found larger $M/L$ ratios for faintest dEs in our sample with respect to the ($M/L$)-$M$ trend of brighter galaxies (see Figure \ref{fig:mtolratio}).

A mixture of formation mechanisms has been considered to model dE formation. In the ``wind model'', dEs are primordial objects that lost their gas in a supernova-driven galactic wind (DS86; Yoshii \& Arimoto 1987: YA87; Chiosi \& Carraro 2002: CC02). In contrast to  massive ellipticals, dwarf ellipticals have very long star formation histories. Since giant galaxies hold on very strongly their gas content, their gas is almost completely converted into stars in a single burst. On the other hand, supernova activities disperse the gas and switch off further star formation in dwarf galaxies. As the gas cools, it sinks back in and again undergoes a new phase of star formation followed by supernova explosions which disperse the gas, and this cycle can be repeated many times (Bender \& Nieto 1990; de Rijcke et al. 2005).

An alternative formation path for dwarf galaxies involves stripping of larger galaxies by gravitational and gas dynamical processes. As galaxies fall in to the cluster they are stripped of much of their gas by ram pressure effects (Gunn \& Gott 1972; van Zee, Skillman \& Haynes 2004), this process can be seen at work in both H$\alpha$ and Ultra-Violet imaging of the Coma cluster (Yagi et al. 2010; Smith et al. 2010). Stars and dark matter can then be removed by interactions with more massive galaxies and the cluster potential, termed ``harassment'' (Moore et al. 1996; Moore et al 1998; Mayer et al. 2001; Mastropietro et al. 2005) or by tidal encounters between galaxies of near-equal mass (Richstone 1976; Aguerri \& Gonz\'alez-Garc\'ia 2009). Both harassment and tidal encounters can create bars (Mastropietro et al. 2005; Aguerri \& Gonz\'alez-Garc\'ia 2009) which could play a part in driving gas not stripped by the ram pressure of the intergalactic medium inwards, triggering star formation. The resulting galaxy is likely to have a higher central surface brightness than its progenitor and would probably resemble a nucleated dE which rotates quite rapidly and still displays some memory of its former state. Although harassment tends to increase internal velocity dispersion, it is unable to disrupt rotational motion and therefore is not obviously reconciled with the low rotational velocities found  by Simien \& Prugniel (2002), Geha et al. (2002, 2003) and MG05.

Another possible way to explain the formation of dwarf galaxies is presented by Kroupa (1998) and Duc et al. (2004). The merging process of two gas-rich disk galaxies results in many massive star clusters which coalesce to produce a small number of dwarf galaxies. These tidally formed dwarf galaxies contain negligible amounts of dark matter and hence cannot explain a significant fraction of the dE population.

Our findings appear to be most consistent with the ``wind model''. In the wind-stripping model, the shallow potential well of dwarf galaxies, with the velocity dispersion of less than 100 km s$^{-1}$  does not allow intense star formation without stripping most of the gas (Schaeffer \& Silk 1988). We found the effective radius-luminosity and Faber-Jackson relation as $R_e \propto L^{0.24}$ and $L \propto \sigma^{2.0}$ in F814W-band, respectively. These two relations are in agreement with the calculations of de Rijcke et al. (2005) based on the winds models of YA87 and CC02. They showed that, although, YA87 models do not predict a tight $L-R_e$ relation, CC02 models can reproduce it very nicely.

Using the accurate measurement of the $M/L$ ratios for the SAURON galaxy sample (Bacon et al. 2001), the FP tilt of early-type galaxies can be attributed to the variation of $M/L$ ratio which is governed by the stellar population and dark matter properties (Cappellari et al. 2006). Moreover, in agreement with Cappellari et al. (2006), considering the resulting $M/L$ ratios based on different star formation models and mass profiles, Allanson et al. (2009) and Grillo \& Gobat (2010) concluded that non-homology is not a major driver of the FP tilt while the mass sensitivity of the $M/L$ ratio is the primary source of FP behaviour in different mass and luminosity regimes. We find that the fainter dEs are bluer, and scatter more about the FP, than the brighter ellipticals. Star formation activity, together with the wind model, can explain how the bluer dwarfs at a given mass have different $M/L$ ratios and thus follow different scaling relations.

\section{Summary}
\label{conclusions}

The results of this study can be summarized as follows:

(i) The FP of bright ellipticals with M$_{814}<-20$ in our sample is fully consistent with the previous studies of the Coma bright galaxies. We found that the scatter about the FP depends at the faint-end luminosity cutoff is significantly increased. We noticed that the FP tilt of faint dEs with M$_{814}>$-20 differs from that of giant ellipticals.

(ii) The FP is not colour dependent which implies that the stellar population plays a insignificant role in the scatter about the FP and changing its tilt. We studied the departure of faint dEs from the FP of bright ellipticals, $\Delta_{FP}$, which is found to be correlated with the luminosity and the galaxy brightness profile. Less concentrated faint dEs, smaller S\'ersic indices, show larger deviation from the FP. We also found a relation between $\Delta_{FP}$ and galaxies central excess light as an imprint of their formation history. In addition, we found a correlation between $\Delta_{FP}$ and the colour of galaxies which indicates that the bluer galaxies in our sample dEs display larger deviation from the FP of the bright ellipticals. On the other hand, we have already shown that in the L-$\sigma$ relation, discussed in paper I, fainter dEs seem to be more massive (e.g. dynamical mass estimated using the internal velocity dispersion) than predicted from their luminosity. Although tidal effects can be important in removing luminous matter from the galaxy, we believe that supernova driven winds are more probably the dominant process.

(iii) Replacing the central velocity dispersion with the S\'ersic index in the FP relation, we studied the photometric plane, PHP, of our sample galaxies. The differences in the coefficients of PHP relation for our dEs and those of previous studies in local elliptical galaxies (GR02), mainly arise from the fainter luminosity coverage of our sample. We noticed that most of the outliers in the projected PHP diagram have internal morphological features or poor S\'ersic fit. Our conclusion is that, while the fundamental plane is less sensitive to detailed morphology of galaxies, the photometric plane is able to effectively differentiate between different morphologies. In other words, the scatter about the FP increases as we extend the sample to fainter dwarf galaxies. In the PHP, however, the scatter is driven by asymmetry and unsmooth galaxy brightness distribution.

(iv) We studied the correlations between the S\'ersic index, $n$, and other photometric or kinematic parameters (i.e. $M_{814}$, $\langle \mu \rangle_e$, $\mu_0$, $\sigma$ and the concentration). The light concentration and S\'{e}rsic index are strongly correlated with the correlation coefficient of 0.95. The process of determining the S\'{e}rsic index, $n$, is model dependent. On the other hand, the concentration parameter is model independent and is determined using a simple photometric analysis. Therefore, based on the $n$-concentration correlation, one is able to economically replace the S\'ersic index, $n$ with the concentration parameter when processing the scaling relations.

The best radius-luminosity relation for our sample dEs is obtained as $R_e \propto L^{0.24}$ which is consistent with the results of de Rijcke et al. (2005) who reported a B-band radius-luminosity power-law slope between 0.28 and 0.55 for a sample of dwarf ellipticals and dwarf spheroidal galaxies.

(v) We found the Kormendy Relation (KR) slopes as 5.23$\pm$0.37 and 5.17$\pm$0.28 for galaxies in the luminosity ranges of $-20<M_{814}<-18$ and $-18<M_{814}<-16$, respectively. This is in agreement with the results of Khosroshahi et al. (2004) who found the KR slope as $\sim$5.2 for dEs of 16 nearby galaxy groups with $-18<M_R<-14$. Furthermore, in agreement with Khosroshahi et al. (2004), we found the steeper linear trends for faint dEs compared to that of the brighter galaxies.

(vi) We obtained the mass-to-light ratio as $M/L \propto \sigma^{-0.25\pm0.33} \propto L^{-0.13\pm0.17}\propto M^{-0.15\pm0.22}$ for our sample dEs indicating that the $M/L$ ratio is almost a flat function of mass, luminosity and velocity dispersion. Our finding agrees well with that of Co09, who reported a $M/L \propto M^{0.09\pm0.06}$ for a sample with nearly the same luminosity range. Galaxies with $M_{814}>-17.5$ seems to have higher $M/L$ ratios with respect to the linear extrapolation of the same associated with brighter galaxies. The variation in $M/L$ ratio is responsible for the deviation of the fainter dEs from the FP. Since, the fainter galaxies in our sample are bluer than the other galaxies and have higher $M/L$ ratios, we attribute their deviation from the FP to their recent star formation activities.

\section*{Acknowledgments}

The authors wish to recognize and acknowledge the very significant cultural role and reverence that the summit of Mauna Kea has always had within the indigenous Hawaiian community.  We are most fortunate to have the opportunity to conduct observations from this mountain. DC acknowledges support from the Science and Technology Facilities Council, under grant PP/E/001149/1. EK would like to acknowledge the financial supports from Sharif University of Technology and IPM.


\bsp

\label{lastpage}

\end{document}